\documentclass[aip,jcp,amsmath,amssymb,reprint,longbibliography,groupedaddress]{revtex4-1}
\usepackage[utf8]{inputenc}
\usepackage{braket}
\usepackage{amsmath}
\usepackage{appendix}
\usepackage{amsfonts}
\usepackage{comment}
\usepackage{cancel}
\usepackage{bbold}
\usepackage{graphicx}
\usepackage{float}
\usepackage{MnSymbol}
\usepackage{mathtools}
\usepackage{tabularx}
\usepackage{array}
\usepackage{amssymb}%
\usepackage{pifont}
\usepackage{bm}
\usepackage{placeins}
\usepackage{MnSymbol}
\usepackage[dvipsnames]{xcolor}

\newcolumntype{Y}{>{\centering\arraybackslash}X}

\usepackage[colorlinks, linkcolor=blue]{hyperref}
\usepackage[nameinlink, capitalise]{cleveref}
\crefname{section}{Sec.}{Sec.}

\newcommand{\nn}{\nonumber} 
\newcommand\numeq[1]%
{\stackrel{\scriptscriptstyle(\mkern-1.5mu#1\mkern-1.5mu)}{=}}

\newcommand{\tcb}{} 

\newcommand\commin[1]{\iffalse #1 \fi}

\newcommand{\mr}{\mathrm}
\newcommand{\mc}{\mathcal}

\newcommand{\eu}[1]{\mathrm{e}^{#1}}

\DeclareMathOperator{\Tr}{Tr}

\newcommand{\e}{\ensuremath{\mathrm{e}}}

\newcommand{\rd}{\ensuremath{\mathrm{d}}}
\newcommand{\thalf}{\ensuremath{\tfrac{1}{2}}}

\newcommand{\sy}{\mathrm{s}}

\DeclareMathOperator{\tr}{tr}

\DeclareMathOperator{\sgn}{sgn}

\hypersetup{
	breaklinks=true,   
	colorlinks=true,   
	pdfusetitle=true,  
}

\begin{document}

\title{Detailed balance in mixed quantum--classical mapping approaches}
\author{Graziano Amati}
\thanks{These authors contributed equally}
\altaffiliation{Present Address: Institute of Physics, University of Freiburg, Freiburg, Germany}
\author{Jonathan R. Mannouch}
\thanks{These authors contributed equally}
\altaffiliation{Present Address: Max Planck Institute for the Structure and Dynamics of Matter, Hamburg, Germany}
\author{Jeremy O. Richardson}
\email{jeremy.richardson@phys.chem.ethz.ch}
\affiliation{Department of Chemistry and Applied Biosciences, ETH Z\"urich, 8093 Z\"urich, Switzerland}

\date{\today}

\begin{abstract}
The violation of detailed balance poses a serious problem for the majority of current quasiclassical methods for simulating nonadiabatic dynamics. In order to analyze the severity of the problem, we predict the long-time limits of the electronic populations according to various quasiclassical mapping approaches, by applying arguments from classical ergodic theory. Our analysis confirms that regions of the mapping space that correspond to negative populations, which most mapping approaches introduce in order to go beyond the Ehrenfest approximation, 
pose the most serious issue for reproducing the correct thermalization behaviour. This is because inverted potentials, which arise from negative electronic populations entering into the nuclear force, can result in trajectories unphysically accelerating off to infinity. 
The recently developed mapping approach to surface hopping (MASH) provides a simple way of avoiding inverted potentials, while retaining an accurate description of the dynamics. We prove that MASH, unlike any other quasiclassical approach, is guaranteed to describe the exact thermalization behaviour of all quantum--classical systems, confirming it as one of the most promising methods for simulating nonadiabatic dynamics in real condensed-phase systems. 
\end{abstract}

\maketitle

\section{Introduction}
Quantum nonadiabatic effects play a crucial role in the description of many relevant physical processes, including radiationless decay, \cite{Matsika2011CI} energy and charge transfer\cite{akimov2014} and coherence control in quantum gates.\cite{magazzu2018exp}
Unfortunately the computational cost of brute-force quantum calculations scales exponentially with 
the number of system degrees of freedom, making it impossible to simulate the vast majority of realistic systems in this way.
An increased interest in the development of quasiclassical methods has therefore arisen, aimed at mapping nonadiabatic dynamics onto effective quasiclassical models that can be simulated as efficiently as classical systems.  
However, improvements in efficiency are often achieved at the expense of accuracy. 
Although rigorous quantum--classical dynamics can be defined in terms of coupled trajectories,\cite{Kapral2015QCL,Martens2016CSH}
we wish to reduce the numerical complexity by evolving trajectories independently.
These quasiclassical methods are often accurate at short times, but can suffer from significant long-time errors. \cite{ultrafast}
In particular, if detailed balance is violated, the final populations will not reflect the correct thermal distribution.

Ehrenfest dynamics is the prototypical quasiclassical nonadiabatic theory that propagates the nuclear dynamics on a mean-field potential energy surface corresponding to a dynamical superposition of electronic states. \cite{tully1998MQC,kirrander2020}
The advantage of this approach is its simplicity, although the accuracy is known to be deficient in many systems.\cite{Stock2005nonadiabatic,Tully2012perspective,identity,spinmap,spinPLDM1,FMOclassical,GQME} This is in part due to the fact that the method drastically violates detailed balance in thermal equilibrium. \cite{Parandekar2006Ehrenfest,ultrafast} 

In order to improve upon the accuracy of Ehrenfest dynamics, a wide array of mean-field mapping approaches have been developed. The main idea is to construct a quasiclassical continuous phase space for the electrons, so that the electronic and nuclear degrees of freedom can be treated on an equal footing.
The most commonly-used mapping for quasiclassical approaches is the Meyer--Miller--Stock--Thoss (MMST) formalism, where the electronic subsystem is mapped onto the single-excitation subspace of a set of harmonic oscillators.
\cite{Meyer1979nonadiabatic,Stock1997mapping,Stock2005nonadiabatic,Sun1998mapping,Wang1999mapping,Kim2008Liouville,Kelly2012mapping,Liu2020linearized} Further improvements in the accuracy of quasiclassical approaches can be achieved by treating the identity operator exactly within the mapping, by representing it by the number one.\cite{identity,FMO,linearized}
A similar improvement is obtained more naturally by departing from the MMST formalism and instead describing the electronic degrees of freedom in terms of a set of spherical spin coordinates, known as spin mapping.\cite{spinmap,multispin} While all these developments often lead to improved dynamics over Ehrenfest theory, they all clearly still violate detailed balance in general.

One important difference between mapping-based approaches and Ehrenfest dynamics is that mapping-space regions corresponding to negative electronic populations are present in the former. 
Although mapping approaches are typically more accurate than Ehrenfest simulations,
these problematic regions can pose a serious issue for the dynamics, 
as the resulting inverted potentials may give rise to an unphysical nuclear force that can cause trajectories to accelerate off to infinity.\cite{Bonella2001mapping1}

In addition to these problems,
there is a conceptual difficulty with using mapping approaches in strongly asymmetric systems in which one initializes the simulation in an excited state and expects the dynamics to thermalize to the ground adiabatic state.
Here, the short-time dynamics must be described by a multi-state nonadiabatic system, whereas the long-time dynamics is an effective one-state problem.
A method that attempts to tackle this issue head-on is the \emph{ellipsoid-mapping} approach,\cite{ellipsoid} which replaces the spherical spin-mapping phase space with an anisotropic ellipsoid geometry.  The shape and orientation of the ellipsoid is dynamically adjusted so as to best represent the effective structure of the local electronic Hamiltonian.
This is achieved in such a way that the dynamics rigorously obey detailed balance by construction.
The approach is however only applicable for computing thermal correlation functions and not able treat the thermalization of systems initialized out of equilibrium, as we wish to do in this paper.  

Another approach designed to go beyond mean-field methods is the symmetric quasiclassical (SQC) approach.\cite{Miller2016Faraday} 
Although the dynamics of SQC are identical to those of the mean-field methods introduced previously, the electronic populations are instead measured by windowing the electronic phase space in such a way that the obtained values are guaranteed to lie in the physical range between zero and one.
An advantage of the method compared to other quasiclassical theories is that SQC is known to obey detailed balance in the limit of zero electron--nuclear coupling.\cite{Miller2015SQC}
However, although it is often an improvement over the original MMST methods, SQC may still lead to inaccurate long-time dynamics in the more general coupled regime, \cite{bellonzi2016}
and is typically of a similar level of accuracy to the spin-mapping methods.\cite{spinmap,multispin}
This is because although SQC avoids measuring negative populations, its trajectories still suffer from their contribution to the nuclear force. 

Fewest switches surface hopping (FSSH)\cite{Tully1990hopping} ensures that the nuclear force is always physical by propagating the nuclei on a single adiabatic surface at any given time. In order to describe nonadiabatic transitions, the active surface is changed stochastically with a probability that mimics the dynamics of the underlying electronic wavefunction. 
\tcb{After a transition, the nuclear momenta are rescaled in the direction of the nonadiabatic coupling vector, consistent with earlier semiclassical scattering theories.\cite{Pechukas1969scattering2,herman1984}}
Remarkably, FSSH has been shown to approximately fulfill detailed balance, even in cases when the electronic and nuclear degrees of freedom are strongly coupled (although rigorously so only in the limit of strong nonadiabatic coupling). \cite{Schmidt2008equilibrium} \tcb{Although much is known about the approximations underlying surface hopping
 (see in particular the work of Subotnik and Kapral)},\cite{Subotnik2013,Kapral2016}
the main problem of FSSH is that the approach still lacks a complete formal derivation. 
\tcb{This means that there is still some disagreement over the correct way of performing certain aspects of the method that could not be derived from first principles, such as the treatment of frustrated hops.} Additionally, the stochastic nature of FSSH means that the electronic wavefunction and the current active surface can become inconsistent during the dynamics, leading to the so-called `inconsistency error' that is known to significantly degrade the accuracy of FSSH observables.\cite{MASH}

It seems that the ultimate quasiclassical approach for accurately describing both the short- and long-time dynamics in nonadiabatic systems is one that combines the best features of SQC and FSSH\@. The mapping approach to surface hopping (MASH) is a newly developed quasiclassical method that achieves just that.\cite{MASH}
MASH consistently windows both the observables and the nuclear force, eliminating the possibility of obtaining negative populations in either. This results in MASH having deterministic dynamics, for which the active surface and the electronic degrees of freedom remain consistent throughout the entire time evolution. The approach is also \tcb{completely} derivable from first principles, leading to a unique prescription for \tcb{frustrated hops} that ensures that the exact short-time dynamics is correctly reproduced. 
Finally, MASH has been tested on a range of commonly-used condensed-phase model systems, where it appears to describe the correct long-time thermalization behaviour.

In this paper, we utilize classical ergodic theory  to predict the long-time thermalization behaviour of a wide range of quasiclassical approaches, including mean-field mapping methods, SQC and MASH\@. By comparing our predictions to those of the expected quantum--classical outcome, we provide a simple and rigorous procedure for benchmarking the long-time dynamics. 
We apply our theory to spin--boson models in challenging parameter regimes and also to an anharmonic model, which together illustrate why negative populations pose a serious problem for reproducing the correct thermalization behaviour for the majority of quasiclassical approaches.  
Despite this, we show that MASH is guaranteed to exactly reproduce the correct thermal populations in the long-time limit for all quantum--classical systems, confirming its potential for being one of the best methods for accurately simulating nonadiabatic dynamics in real condensed-phase systems. 

\section{Theory}\label{sec:theory}
In the present work, we focus on the dynamics of a quantum subsystem consisting of $N$ states, $\ket{k}$, coupled to a classical environment, in this case a heat bath.
The general form of the Hamiltonian is
\begin{equation}\label{eq:H}
\hat H(x,p) = \frac{p^{2}}{2m}+U(x) +\hat{V}(x) ,
\end{equation}
where $(x,p)$ are the multidimensional classical phase-space variables of the environment. 
$U(x)$ and $\hat{V}(x)$ are state-independent and state-dependent potentials respectively; the latter is a traceless $N\times N$ matrix in the \tcb{diabatic} basis\tcb{, $\{\ket{k}\}_{k=1}^N$}. 
Let us remark that it is valid to treat the bath classically provided that the energy associated with the thermal fluctuations of the environmental modes is large compared to their zero-point energies, although not necessarily large compared to the energy scales of the quantum subsystem. 
In the framework of atomistic systems in the condensed phase, the subsystem and the environment typically refer to electronic and nuclear degrees of freedom respectively, although this particular situation is not strictly required in the following.

The aim of quasiclassical approaches is to accurately predict the evolution of quantum correlation functions, with a computational cost comparable to classical simulations. 
Approximate quasiclassical dynamics associated with \cref{eq:H} can be derived by first taking the partial Wigner transform of the quantum Hamiltonian and then taking the classical limit of the bath. \cite{leaf1968,sergi2004}  
The quantum subsystem is mapped onto a continuous phase-space representation, via Cartesian mapping variables, $X,P\in \mathbb R^N$. \cite{Stock2005nonadiabatic,NRPMDChapter}
This leads to a representation of the full system in terms of the phase-space points, $\Gamma=\{X,P,x,p\}$. The Hamiltonian [Eq.~(\ref{eq:H})] is then mapped onto a phase-space function, $\mathcal{H}(\Gamma)=p^2/2m+U(x)+\mathcal{V}(x,X,P)$, from which 
the dynamics 
are obtained. 

Quasiclassical approaches derived in this way result in equations of motion of the form
\begin{subequations}\label{eq:eom}
\begin{align}
\dot{X}_{k}&=\sum_{k'}\braket{k|\hat{V}(x)|k'}P_{k'}, \label{eq:eomX}\\
\dot{P}_{k}&=-\sum_{k'}\braket{k|\hat{V}(x)|k'}X_{k'},\label{eq:eomP}\\
\dot{x}_{j}&=\frac{p_{j}}{m_{j}}, \label{eq:eomx}\\
\dot{p}_{j}&=-\frac{\partial U(x)}{\partial x_{j}}+\mathcal{F}_{j}(x,X,P) . \label{eq:eomF}
\end{align}
\end{subequations}
Here we have assumed that the matrix elements $\braket{k|\hat{V}(x)|k'}$ are real.
The first two lines are equivalent to the Schr\"odinger equation for the real and imaginary parts of the electronic wavefunction.
The last two lines are Newton's equations of motion for the bath modes.
The expressions for the state-dependent potential, $\mathcal{V}(x,X,P)$ and nuclear force, $\mathcal{F}_{j}(x,X,P)$, depend on the specific method considered. Apart from the total energy, $\mathcal{H}(\Gamma)$, these equations of motion additionally conserve the norm of the Cartesian mapping variables, $r=\thalf\sum_{k=1}^{N}(X_{k}^{2}+P_{k}^{2})$. 

From these dynamics, quantum correlation functions, $\Tr[\hat{\rho}_0\hat{A}\hat{B}(t)]$, are then approximated by
\begin{equation}\label{eq:CAB}
\mc C_{AB}(t) = \int\mr d  \Gamma\,\rho_0(\Gamma)A(\Gamma)B(\Gamma_t),
\end{equation}
where the observable operators are mapped onto phase-space functions [i.e., $\hat{A}(x,p)\mapsto A(\Gamma)$], the specifics of which are method dependent. 
Note that in some approaches, the quasiclassical representations of the initial and time-evolved operators can differ, even if quantum mechanically the operators are identical (due to different projections of electronic operators \cite{Kelly2012mapping,identity} or due to the inclusion of weighting factors into correlation functions \cite{MASH}).

A well known limitation of the majority of quasiclassical methods is their inability to obey detailed balance and recover the correct long-time populations.\cite{Parandekar2006Ehrenfest,Mueller1998mapping,ultrafast,ellipsoid}
This issue results in dynamical correlation functions relaxing to incorrect long-time limits. As we will see, the way in which the correlation function operators and the state-dependent force are represented by phase-space functions can lead to significant differences in the thermalization properties of different methods. 

\subsection{Mean-field approaches} \label{subsec:MF}
For mean-field approaches, the state-dependent potential is approximated by its expectation value with respect to the Cartesian mapping variables. 
This gives rise to dynamics that correspond to the nuclei being propagated on a weighted average of the potential energy surfaces
\begin{subequations}
\begin{align}
\mathcal{V}(x,X,P)&=\frac{1}{2}\sum_{kk'}\braket{k|\hat{V}(x)|k'}(X_{k}X_{k'}+P_{k}P_{k'}) , \label{eq:pot_mf} \\
\mathcal{F}_{j}(x,X,P)&=-\frac{1}{2}\sum_{kk'}\frac{\partial\braket{k|\hat{V}(x)|k'}}{\partial x_{j}}(X_{k}X_{k'}+P_{k}P_{k'}) . \label{eq:mean-field_force}
\end{align}
\end{subequations}
Let us remark that Eqs.~\eqref{eq:eom} are simply Hamilton's equations of motion given the mean-field form of the state-dependent potential [\cref{eq:pot_mf}].

The prototypical mean-field approach is Ehrenfest dynamics, for which both the nuclear force and the electronic observables are given by expectation values associated with the normalized time-evolved electronic wavefunction (here expressed in terms of mapping variables). 
Improved mean-field methods have been obtained by mapping the electronic subsystem onto a fictitious system containing continuous degrees of freedom, $(X,P)$, such that the quantum and nuclear subsystems can be treated on an equal footing. 
Whereas in Ehrenfest theory, $X$ and $P$ simply correspond to the real and imaginary parts of the electronic wavefunction,
in the mapping approaches, they span a space in which representations of the wavefunction are constructed.
While the mean-field expression for the state-dependent nuclear force remains the same [\cref{eq:mean-field_force}], the phase-space representations of the observable operators generally depend on the mapping space used. 

In the case of the Meyer--Miller--Stock--Thoss (MMST) mapping, \cite{Meyer1979nonadiabatic,Stock1997mapping} electronic states are mapped onto the single-excitation subspace of a set of $N$ harmonic oscillators.
This leads to (at least) two possible representations of the electronic operators in terms of phase-space functions.\cite{Kelly2012mapping,identity} 
In the Wigner approach, the phase-space functions are obtained from the Wigner transform of the associated operators in the mapped harmonic-oscillator system, analogous to the representation of the potential energy operator used in \cref{eq:pot_mf}. 
In the singly-excited oscillator (SEO) approach,\cite{Stock1997mapping} projection operators onto the physically relevant single-excitation subspace are added, which results in an additional factor of $\phi(r)=16\eu{-2r}$ appearing from the Wigner transform. 
At least one factor of $\phi(r)$ is required for the integral over mapping variables to converge.
These exponential factors are normally incorporated into the definition of the initial density, $\rho_{0}(\Gamma)$, as they define the distribution from which the Cartesian mapping variables are initially sampled. We will call this contribution $\rho_{0,\mr{s}}$.

Both of these approaches lead to the representation of the identity operator, $\mathcal{I}_{\mr{s}}$, being a function of the mapping variable norm, $r$. It was however previously observed that representing the identity operator with the number one (which can be thought of as the exact mapping for this operator) can lead to more accurate results. \cite{identity,FMO,linearized} We refer to MMST methods that represent the identity operator in this way as `unity approaches'.

Another variant of MMST mapping utilizes so-called focused initial conditions when the dynamics are initialized in an electronic population,\cite{Kim2008Liouville} by initially sampling the mapping variables from the regions that satisfy $A(\Gamma)=1$.
\tcb{These thus have delta functions for their initial $r$-distributions.  All Ehrenfest trajectories have $r=1$ due to the absence of electronic zero-point energy, whereas in focused MMST methods, which utilize the energy levels of a harmonic oscillator, the populated state contributes $\frac{3}{2}$ and the $N-1$ unpopulated states contribute $\frac{1}{2}$ each, leading to $r=\frac{3}{2}+\frac{1}{2}(N-1)=\frac{1}{2}(N+2)$.}

For spin mapping,\cite{spinmap,multispin} the electronic states are described by a set of spin variables, which results in the following advantageous features over MMST mapping. First, the associated mapping space is isomorphic with the original quantum subspace, such that the use of projection operators is no longer required and the representation of observable operators is therefore unique. Second, the correspondence between the identity and the number one arises naturally from the underlying theory and does not need to be imposed, as in the case of the unity approaches.
The algorithm is practically identical to the MMST-focused method except that the hypersphere radius is $r=\sqrt{N+1}$.
Spin mapping in its original formulation is a linearized quasiclassical method. \cite{spinmap,multispin}
Spin-PLDM is a partially linearized extension \cite{spinPLDM1,spinPLDM2} developed within the framework of the partially linearized density matrix (PLDM) approach. \cite{Huo2011densitymatrix,Hsieh2012FBTS}
The equations of motion are similar to those of the fully-linearized mapping methods except that two sets of mapping variables are employed to describe the forward and backward propagation separately.

\Cref{tab:comparison} summarizes the different ways in which observable operators are represented by the methods studied in this work. 
We remark that double-SEO is also known as the linearized semiclassical initial value representation (LSC-IVR),\cite{Sun1998mapping,Wang1999mapping} while single-SEO is often referred to as the Poisson bracket mapping equation (PBME).\cite{Kim2020mapping,Kelly2012mapping}
Most cases are \tcb{easily obtained by rewriting the expressions from the original papers\cite{identity,spinmap,multispin} in terms of $r$}.
The more obscure spin-PLDM expression for $\rho_{0,\mr s}$ given in \cref{tab:comparison} has first been derived in Ref.~\onlinecite{runesonPhD} by integrating out all other spin degrees of freedom apart from the centroid. Note that this particular expression is only valid for two-state systems.

One of the major problems of most mean-field approaches is the presence of negative populations, which can lead to both unphysical values for the long-time limit of correlation functions and unphysical dynamics corresponding to propagating the nuclei on inverted potentials. 
In the following, we introduce two approaches that are designed to alleviate the problem of negative populations, namely the symmetric quasiclassical theory (SQC) and the mapping approach to surface hopping (MASH).

\begin{table*}
\caption{A comparison between the different quasiclassical methods studied in this work. 
For each method, the initial mapping-variable distribution, $\rho_{0,\mr{s}}$ and the identity operator representation, $\mathcal{I}_{\mr{s}}$ are given. 
Here, $\phi(r)=16\eu{-2r}$.
\tcb{
Derivations of the radial distributions and representations of the identity for different mean-field approaches can be found in Refs.~\onlinecite{identity,spinmap}.}
Additionally, the accuracy of the thermalization behaviour of each method 
in the high-temperature ($\beta\to 0$), low-temperature ($\beta\to \infty$) and weak electron-nuclear coupling ($\alpha\to 0$) limits is indicated, 
valid for any two-state system. We mark with ticks all entries that agree with the correct quantum--classical result, along with any erroneous multiplicative factors for the $\beta\to 0$ and $\beta\to \infty$ limits. For example, Ehrenfest predicts the long-time adiabatic population difference a factor of 3 too small when $\beta\rightarrow0$ but is correct when $\beta\rightarrow\infty$. 
Further details on the representation of the operators in different methods are given in \cref{sec:analysis}.
}\label{tab:comparison}
\begin{ruledtabular}\begin{tabular}{lccccc}
Method & $r^{N-1}\rho_{0,\mr s}$ & $\mathcal{I}_{\mr s}$ & $\beta\rightarrow 0\footnotemark[1]$ & $\beta\rightarrow \infty\footnotemark[1]$ & $\alpha\rightarrow0\footnotemark[1]$ \\ [1.5pt]
\hline
MASH\footnotemark[1]  & $\mathcal{W}_{AB}(\bm{S}^{\mr{ad}})$ & 1 & \textcolor{teal}{\textbf{$\checkmark$}} & \textcolor{teal}{\textbf{$\checkmark$}} & \textcolor{teal}{\textbf{$\checkmark$}}  \\ [1.5pt]
SQC\footnotemark[1] & $h(2-r)$ & $h(|rS^{\text{ad}}_{z}|-2+r)$ & \textcolor{teal}{\textbf{$\checkmark$}} & \textcolor{teal}{\textbf{$\checkmark$}} & \textcolor{teal}{\textbf{$\checkmark$}}  \\ [1.5pt]
Spin mapping& $\delta(r-\sqrt{N+1})$  & $1$ & \textcolor{teal}{\textbf{$\checkmark$}} & $\times\sqrt{3}$ & $\times$ \\ [1.5pt] 
Spin-PLDM\footnotemark[1] & $\tfrac{4}{3}rh(\sqrt{3}-r)$ & $1$ & \textcolor{teal}{\textbf{$\checkmark$}} & $\times\tfrac{4\sqrt{3}}{3}$ & $\times$  \\ [1.5pt]
Ehrenfest & $\delta(r-1)$ & 1 & $\times\tfrac{1}{3}$ & \textcolor{teal}{\textbf{$\checkmark$}} & $\times$  \\ [1.5pt]
MMST-focused\footnotemark[2] & $\delta(r-\tfrac{1}{2}(N+2))$ &1 & $\times\tfrac{4}{3}$ & $\times 2$ & $\times$ \\ [1.5pt]
Single-Wigner\footnotemark[2] & $\tfrac{1}{2}r^{N-1}\phi(r)$ & $r-1$ & \textcolor{teal}{\textbf{$\checkmark$}} & \textcolor{teal}{\textbf{$\checkmark$}} & $\times$  \\ [1.5pt]
Double-SEO\footnotemark[2] & $\tfrac{1}{2}r^{N-1}\phi^{2}(r)$ & $r-\thalf$ & $\times\thalf$ & \textcolor{teal}{\textbf{$\checkmark$}} & $\times$  \\ [1.5pt]
Single-SEO\footnotemark[2] & $\tfrac{1}{2}r^{N-1}\phi(r)$ & $r-\thalf$ & $\times\tfrac{3}{2}$ & $\times 2$ & $\times$ \\ [1.5pt]
Double-unity\footnotemark[2] &$\tfrac{1}{2}r^{N-1}\phi^{2}(r)$ & $1$ &  \textcolor{teal}{\textbf{$\checkmark$}}  & $\times 4$ & $\times$  \\ [1.5pt]
Single-unity\footnotemark[2] & $\tfrac{1}{2}r^{N-1}\phi(r)$ & $1$ & \textcolor{teal}{\textbf{$\checkmark$}} & $\times 2$ & $\times$  \\ [1.5pt]
\end{tabular}\end{ruledtabular}
\footnotetext[1]{Formulas and results for the case of $N=2$ only}
\footnotetext[2]{Mean-field MMST mapping method}
\end{table*}

\subsection{Symmetric quasiclassical windowing (SQC)}\label{subsec:SQC1}
The symmetric quasiclassical approach (SQC) 
attempts to improve upon the standard MMST mapping approaches
by measuring the electronic populations with `windows' that guarantee that population observables always physically lie between zero and one. 
These windows are usually defined in terms of the action-angle variables ($n_{k}\ge0$ and $q_{k}$ respectively), which are related to the Cartesian mapping variables introduced in \cref{sec:theory} as follows: \cite{Cotton2013mapping}
\begin{subequations}
\label{eq:AA}
\begin{align}
X_{k}&=\sqrt{2(n_{k}+\gamma)}\cos{q_{k}}, \\
P_{k}&=-\sqrt{2(n_{k}+\gamma)}\sin{q_{k}} .
\end{align}
\end{subequations}
The most successful choice for the windowing functions are the so-called `triangular windows', which for state $k$ take the form \cite{Cotton2019FMO}
\begin{equation}
\label{eq:window}
\begin{split}
W_{k}(n)&=(2-\gamma-n_{k})^{2-N}h(n_{k}+\gamma-1)h(2-\gamma-n_{k}) \\
&\times\prod_{k'\neq k}h(2-2\gamma-n_{k}-n_{k'}) ,
\end{split}
\end{equation}
where $h(\cdot)$ denotes the Heaviside step function, the zero-point energy parameter is set to 
$\gamma=1/3$,
and the angle variables, $q_{k}$, are uniformly sampled independently from the interval $[0,2\pi)$.
The success of this windowing scheme is due to the fact that the windows associated with different electronic states touch, allowing the approach to correctly describe population transfer even between weakly-coupled states. \cite{Cotton2016SQC} 
However, because the windowing functions given by \cref{eq:window} do not fill the entire mapping space, the sum of the populations is not a constant of motion. 
Therefore, all SQC observables must be renormalized to obtain a total population of one whenever they are measured. 

While this scheme guarantees that the population observables always return physical values, the approach still suffers from the effects of inverted potentials, because the dynamics evolve via the same Hamiltonian as for the mean-field methods. 
This means that while it has been shown that SQC correctly thermalizes in the limit of weak system-bath coupling, \cite{Miller2015SQC} it is expected that this is in general not true in other parameter regimes, particularly for strongly-coupled and anharmonic systems.
One suggestion to alleviate the problem of inverted potentials is to introduce a state-dependent zero-point energy parameter, $\gamma_{k}$, into the definition of the state-dependent nuclear force [\cref{eq:mean-field_force}], and adjust their values independently for each trajectory so that the contribution to the force from each electronic population is correctly reproduced.\cite{Cotton2019SQC} 
On the one hand, this strategy can reduce the likelihood of trajectories unphysically accelerating off to infinity (at least at short times).
On the other hand, this means that the values of $\gamma_{k}$ used for measuring the observables and performing the dynamics are inconsistent. 

\subsection{Mapping approach to surface hopping (MASH)}\label{subsec:MASH}
The mapping approach to surface hopping (MASH) \cite{MASH} is a new nonadiabatic trajectory approach that offers the best of both worlds between FSSH and quasiclassical mapping dynamics. MASH can be thought of as going beyond SQC by windowing both the observables and the nuclear force, thus solving the problem of inverted potentials completely. 
Windowing forces in this way introduces hops similar to those of FSSH,
although the inconsistency error of FSSH is not present in MASH\@.
These appealing features make MASH one of the most promising approaches for performing accurate nonadiabatic dynamics within realistic ab initio simulations of molecules at a relatively low computational cost.

MASH was originally derived for systems involving two electronic states  using the spin-mapping variables in the adiabatic basis, given by  
\begin{subequations}
\label{eq:spinmap_var}
\begin{align}
rS_{x}^{\mr{ad}}&=X_{+}X_{-}+P_{+}P_{-} , \\
rS_{y}^{\mr{ad}}&=X_{+}P_{-}-X_{-}P_{+} , \\
rS_{z}^{\mr{ad}}&=\tfrac{1}{2}(X^{2}_{+}+P^{2}_{+}-X^{2}_{-}-P^{2}_{-}) ,
\end{align}
\end{subequations}
where $\pm$ refer to the upper and lower adiabatic states. \tcb{Like FSSH, MASH works within the kinematic picture, where the nuclear momentum retains the intuitive meaning of mass times velocity, unlike the canonical momentum associated with the Hamiltonian in the adiabatic representation.\cite{Cotton2017mapping} Thus the MASH expression for the energy takes the same form as in \cref{eq:H}, except that the state-dependent potential is now given in the adiabatic basis.} 
In terms of these mapping variables, the MASH representation of the state-dependent potential and the nuclear force operator are given by\cite{MASH}
\begin{subequations}\label{eq:pot/force_MASH}
\begin{align}
\mathcal{V}(x,S^{\mr{ad}})&=V_{z}^{\mr{ad}}(x)\,\sgn(S_{z}^{\mr{ad}}) , \label{eq:pot} \\
\mathcal{F}_{j}(x,S^{\mr{ad}})&=-\frac{\partial V^{\mr{ad}}_{z}(x)}{\partial x_{j}}\,\sgn(S_{z}^{\mr{ad}}) \nn \\
&\quad +4V^{\mr{ad}}_{z}(x)d_{j}(x)S^{\mr{ad}}_{x}\delta(S^{\mr{ad}}_{z}) , \label{eq:force_MASH}
\end{align}
\end{subequations}
where $\sgn(\cdot)$ returns the sign of its argument, the adiabatic potential energy surfaces are $V_\pm(x) = U(x) \pm V_{z}^{\mr{ad}}(x)$, $d_{j}(x)$ is the nonadiabatic coupling vector\cite{ConicalIntersections1} and $\delta(\cdot)$ is the Dirac delta function. 
From \cref{eq:pot} it can be seen that the adiabatic windows used in MASH 
correspond to the spin-hemispheres. These windows, 
like the triangular windows of SQC, touch at the equator,
but
unlike SQC windows, they fill the whole mapping space, which is necessary in MASH to guarantee that the force is well defined throughout the entire time-propagation. 
The last term on the right-hand side of \cref{eq:force_MASH} ensures energy conservation by applying an impulse at the equator. As MASH is guaranteed to exactly reproduce the short-time dynamics of the quantum--classical Liouville equation,\cite{Kapral2015QCL} this term therefore constitutes a unique and correct prescription for the momentum rescaling and frustrated hops, which is a feature that is lacking in almost all previous surface-hopping approaches.
In practice the algorithm is simply a deterministic version of FSSH with momentum rescalings at each attempted hop. 
Finally, the initial mapping-variable distribution, $\rho_{0,\mr{s}}=\mathcal{W}_{AB}(\bm{S}^{\mr{ad}})$, is determined so that the MASH windows exactly reproduce the Rabi oscillations of a bare electronic system.\cite{MASH}

\tcb{Because of its use of the kinematic picture,} the equations of motion associated with MASH are not obviously generated by a Hamiltonian. 
This means that $\mathcal{H}(\Gamma)$ with the MASH representation of the potential [\cref{eq:pot}] is technically a conserved energy rather than the generator of the quasiclassical dynamics.
This raises the question of whether applying classical ergodic theory to predict the long-time limit of the MASH correlation functions is justified. 
Despite this, the method does possess many of the features associated with Hamiltonian dynamics, such as the conservation of phase-space volume,  
suggesting that the use of classical ergodic theory may still be applicable. 
In \cref{subsec:exp_cross}, we will show numerically that the long-time limits of MASH correlation functions do agree with the predictions of classical ergodic theory, which justifies our use of it for MASH. 

\subsection{Ergodic Theory}\label{subsec:erg}
Quasiclassical approaches that are both deterministic and based on independent trajectories can be analyzed within the framework of classical ergodic theory.
In this section, we expand this idea to derive a general expression for the long-time limits of correlation functions, \cref{eq:long_time_gen}.
As previously discussed, the equations of motion  [\cref{eq:eom}] conserve the norm of the Cartesian mapping variables, $r=\tfrac{1}{2}\sum_{k=1}^{N}(X_{k}^{2}+P_{k}^{2})$.
We cannot therefore simply apply classical ergodic theory directly to our problem. 
In fact, according to the so-called \emph{ergodic hierarchy}, \cite{berkovitz2006} a necessary requirement for the dynamics to exhibit mixing behavior on a given manifold of the phase space is the condition of ergodicity, i.e., only the Hamiltonian, $\mc H(\Gamma)$, is allowed to be conserved on that manifold. 
Thus, in order to isolate the additional conserved variable, $r$, we define a set of hyperspherical coordinates for the mapping variables, $(X,P)\mapsto(r,\Omega)$, where $\Omega$ denotes the solid angle.
By introducing $\tilde{\Gamma} = (\Omega, x, p)$,
we can rewrite \cref{eq:CAB} as
\begin{subequations}\label{eq:CAB2} 
\begin{align}
\mc C_{AB}(t) &= 
\int_0^{\infty}\mathrm d r\,  r^{N-1}\mc C_{AB}^{(r)}(t),\label{eq:CAB21} \\
\mc C_{AB}^{(r)}(t) &= \int \mr d \tilde{\Gamma} \,\rho_{0}(r,\tilde{\Gamma} ) A(r, \tilde{\Gamma} )B(r,\tilde{\Gamma} _t).\label{eq:CAB22}
\end{align}
\end{subequations}

To evaluate the long-time limit of \cref{eq:CAB2}, we define $\rho^{(r)}_{t,0}(\tilde{\Gamma} '|\tilde{\Gamma} ) $ as the conditional probability of occupying the state $\tilde{\Gamma} '$ at time $t$, given that the dynamics propagate on the hypersurface of constant $r$ and that the system was initialized in state $\tilde{\Gamma}$.
\Cref{eq:CAB22} is then expanded as
\begin{align}
\mc C_{AB}^{(r)}(t) &= \int \mr d \tilde{\Gamma} \, \rho_{0}(r,\tilde{\Gamma} ) A(r,\tilde{\Gamma} ) \int\mr d \tilde{\Gamma} '\,  \rho^{(r)}_{t,0}(\tilde{\Gamma} '|\tilde{\Gamma} ) B(r,\tilde{\Gamma}').\label{eq:CABR}
\end{align}
We will assume that the dynamics fulfill the \textit{strong mixing condition}\cite{hawkins2021,berkovitz2006}
\begin{align}\label{eq:mixing}
\begin{split}
\lim_{t\to\infty} \rho^{(r)}_{t,0}(\tilde{\Gamma} '|\tilde{\Gamma} ) 
&=\rho_\mathrm{eq}(r,\tilde{\Gamma} ')= \frac{\e^{-\beta \mathcal{H}(r,\tilde{\Gamma} ')}}{\mathcal{Z}(r)} ,
\end{split}
\end{align}
where 
$\mathcal{Z}(r) = \int \mr d \tilde{\Gamma} \, \e^{-\beta \mathcal{H}(r,\tilde{\Gamma} )}$.
The equilibrium canonical distribution in \cref{eq:mixing} is the long-time distribution expected for a classical ergodic system with finite temporal correlations. \cite{evans2009}
The inverse temperature, $\beta$, in \cref{eq:mixing} is obtained using the assumption of equipartition of energy in thermal equilibrium.
\Cref{eq:mixing} implies that as $t\to\infty$, the probability of reaching any point $\tilde\Gamma'$ on a given manifold at constant $r$ does not depend on the initial conditions. 
This is expected to be valid in most models that include a large number of bath degrees of freedom that couple 
to the relevant subsystem.
With \cref{eq:CABR,eq:mixing}, we obtain the following expression for the long-time limit of quasiclassical correlation functions:
\begin{subequations}\label{eq:long_time_gen}
\begin{align}
\mc C_{AB}(t\rightarrow\infty) 
& = \int_0^{\infty}\mathrm d r\,r^{N-1} \langle  A\rangle_{0}^{(r)}\langle B\rangle_{\mr{eq}}^{(r)},\label{eq:long_time} \\
 \langle  A\rangle_{0}^{(r)} &= \int \mr d \tilde{\Gamma} \,\rho_{0}(r,\tilde{\Gamma} ) {A}( r,\tilde{\Gamma} ),\label{eq:Ai_ave}\\
\langle B\rangle_{\mr{eq}}^{(r)} &= \frac 1{\mathcal{Z}(r)}\int\mathrm d \tilde{\Gamma} \,\e^{-\beta \mathcal{H}(r,\tilde{\Gamma} )} {B}(r,\tilde{\Gamma} ).\label{eq:B2_ave}
\end{align}
\end{subequations}

The correct long-time limit expected for the quantum--classical correlation function \tcb{as $\hbar\rightarrow 0$} is however given by
\begin{subequations}\label{eq:benchmark_gen}
\begin{align}
\mc C_{AB}^{\mathrm{QC}}(t\rightarrow\infty)&=\braket{A}^{\mathrm{QC}}_{0}\braket{B}^{\mathrm{QC}}_{\mathrm{eq}} , \\
\braket{A}^{\mathrm{QC}}_{0}&=\int\rd x\rd p\,\tr_{\mr s}[\hat{\rho}_{0}(x,p)\hat{A}(x,p)] , \\
\braket{B}^{\mathrm{QC}}_{\mathrm{eq}}&=\frac{1}{Z_{\mathrm{QC}}}\int\rd x\rd p\,\tr_{\mr s}[\eu{-\beta \hat{H}(x,p)}\hat{B}(x,p)] , \label{eq:boltz_qc}
\end{align}
\end{subequations}
where $\tr_\sy[\cdot]$ denotes the partial trace with respect to the subsystem and $Z_{\mathrm{QC}}=\int\rd x\rd p\,\tr_\sy[\eu{-\beta\hat{H}(x,p)}]$. \cite{mauri1993} 
This provides the benchmark result against which we will test the various quasiclassical predictions. \tcb{While higher-order contributions in $\hbar$ to the quantum--classical Boltzmann operator do in principle exist,\cite{nielsen2001} we do not consider them here, as almost all quasiclassical approaches cannot even reproduce the dominant zeroth-order term correctly in all cases.
Additionally we show in \cref{app:ham_rep} that \cref{eq:benchmark_gen} is independent of the Hamiltonian representation, as long as the $\hbar\rightarrow0$ limit is taken.}

The main difficulty for quasiclassical approaches in reproducing this long-time limit is found in the term given by \cref{eq:boltz_qc}. 
This is because the majority of mappings implemented in quasiclassical approaches can only at most reproduce the correct trace relations for a product of two operators, $\int\rd x\rd p\,\tr_\sy[\hat{H}(x,p)\hat{B}(x,p)]=\int \rd\Gamma \,\mathcal{H}(\Gamma){B}(\Gamma)$, and will therefore not be able to correctly describe all the terms arising from the Taylor expansion of the Boltzmann operator in \cref{eq:boltz_qc}. 
In order to better understand the relative accuracy of different quasiclassical approaches in reproducing the correct long-time limit of correlation functions, we apply this analysis to the specific case of two-state systems coupled to a heat bath.

\section{Application to two-level systems}\label{sec:two_level}
The arguments derived so far hold for a subsystem consisting of an arbitrary number of quantum levels. 
For the sake of simplicity, here and in the following we apply our analysis to two-level quantum systems, although the majority of our arguments apply equally well to multi-state problems. 
In this case, the state-dependent potential in \cref{eq:H} can be written as
\begin{equation}\label{eq:V}
\begin{split}
\hat{V}(x)&=\Delta(x)\hat{\sigma}_{x}+\kappa(x)\hat{\sigma}_{z} , \\
&\equiv V_{z}^{\text{ad}}(x)\hat{\sigma}^{\text{ad}}_{z}(x),
\end{split}
\end{equation}
where $\Delta(x)$ and $\kappa(x)$ denote diabatic Hamiltonian parameters and $\hat\sigma_i$,  for $i=x,y,z$, are the Pauli operators in the diabatic basis. These, together with the $2\times 2$ identity, $\hat{\mathcal{I}}_{\text{s}}$, form a complete set of Hermitian operators for the two-level system.
The $\hat{\sigma}^{\text{ad}}_i(x)$ operators are the Pauli matrices in the adiabatic basis, related to the diabatic $\hat\sigma_i$ operators by a linear transformation. \cite{MASH}
Finally, $V_{z}^{\text{ad}}(x)=\sqrt{\Delta(x)^{2}+\kappa(x)^{2}}$ denotes half the energy gap in the adiabatic basis.

We test our theoretical predictions on two nonadiabatic models in which the state-dependent potential depends on a one-dimensional reaction coordinate, $x_{\mr c}$.  
This convenient choice implies that the nuclear phase-space integrals in \cref{eq:long_time_gen} become one-dimensional, simplifying the interpretation of our results. 
An $f$-dimensional secondary bath, with coordinates $x_1,\dots,x_f$, is introduced to provide friction on the reaction coordinate, such that the dynamics thermalize in the long-time limit.  
The secondary bath interacts with the subsystem only via the reaction coordinate and can therefore be easily integrated out of the long-time limit expressions. \cite{ellipsoid}
The two contributions to the potential in \cref{eq:H} are then expressed as
\begin{subequations}\label{eq:gen_pot}
\begin{align}
U(x) &= \frac{1}{2}\sum_{j=1}^{f}\omega_{j}^{2}\left(x_{j}+\frac{c_{j}x_{\mr c}}{\omega_{j}^{2}}\right)^{2}+ U_{\mathrm{RC}}(x_{\mr c}),\\
\hat{V}(x) &= \hat{V}_{\mathrm{RC}}(x_{\mr c}),
\end{align}
\end{subequations}
where $U_{\mathrm{RC}}(x_{\mr c})$ and $\hat{V}_{\mathrm{RC}}(x_{\mr c})$ depend on the specific model considered and $m=1$ for all degrees of freedom, which corresponds to working in mass-weighted coordinates.
We also choose a purely Ohmic spectral density for the secondary bath
\begin{equation}\label{eq:pure_ohm}
J(\omega)=\eta\omega.
\end{equation}
The use of an Ohmic spectral density means that the dynamical simulations can be easily performed by evolving the reaction coordinate using Langevin equations of motion, which implicitly describe the effects of the secondary bath. \cite{TuckermanBook}

In the following, we focus on the long-time dynamics of  $\hat{B}=\hat{\sigma}^{\text{ad}}_{z}(x_{\mr c})$, which corresponds to the population difference between the two adiabatic states. Additionally, we consider a factorized initial condition, where the electronic subsystem is initialized in $\hat{A}=\tfrac{1}{2}\hat{\mathcal{I}}_{\mr{s}}$ and the phase-space variables associated with the reaction coordinate are sampled from an initial Gaussian distribution
\begin{equation}
\label{eq:nuc_dist}
\rho_{\mr{b}}(x_{\mr c},p_{\mr c})=\frac{\beta\Omega}{2\pi}\e^{-\beta(\frac{1}{2}p_{\mr c}^2+\frac{1}{2}\Omega^{2} x_{\mr c}^{2})},
\end{equation}
where $\Omega$ defines its physical width.
We call this correlation function $\mathcal{C}_{\mathcal{I}z}(t)$.
Note that other correlation functions with observables orthogonal to $\hat{\sigma}^{\text{ad}}_{z}(x_{\mr c})$ [e.g., the coherences $\hat{\sigma}^{\text{ad}}_x(x_{\mr c})$ and $\hat{\sigma}^{\text{ad}}_{y}(x_{\mr c})$] have \tcb{to zeroth-order in $\hbar$} a zero expectation value in the long-time limit by symmetry.
This implies that if the long-time limit of $\mathcal{C}_{\mathcal{I}z}(t)$ is captured correctly, so will all other correlation functions, as long as they are constructed from the appropriate linear combinations of the adiabatic observables.

\subsection{Spin--boson model}\label{subsec:sys_bath}
The spin--boson model \cite{Leggett1987spinboson} is commonly used to describe charge transfer in the condensed phase. 
Despite its apparent simplicity, it provides a stringent test for quasiclassical methods, especially in strongly asymmetric cases due to the problem of negative populations.
In the reaction-coordinate picture, the potentials associated with the spin--boson model are\cite{Thoss2001hybrid} 
\begin{subequations}
\begin{align}
U_{\mathrm{RC}}(x_{\mr c})&=\tfrac{1}{2}\Omega^{2}x_{\mr c}^{2} , \\
\hat{V}_{\mathrm{RC}}(x_{\mr c})&=\Delta\hat{\sigma}_{x}+(\varepsilon+\alpha x_{\mr c})\hat{\sigma}_{z} ,
\end{align}
\end{subequations}
where $\Omega$ is the frequency of the reaction coordinate (which we also choose to coincide with the width of the initial Gaussian distribution [\cref{eq:nuc_dist}]), $\Delta$ is the coupling between the two diabatic states and $\alpha$ is the system--bath coupling strength.
We employ reduced units where mass and $\hbar$ are 1.

\begin{figure}
\centering
\includegraphics[width=3.25in]{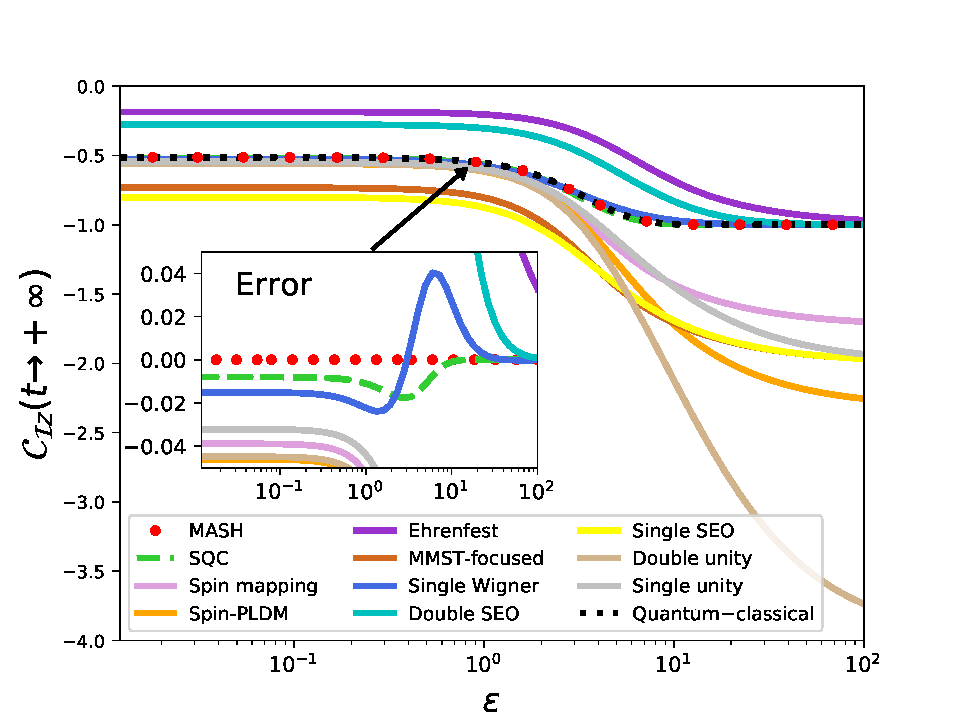}\caption{The long-time limits of the adiabatic population difference as a function of the energy bias, $\varepsilon$, for the spin--boson model introduced in \cref{subsec:sys_bath}. The  other parameters are fixed to $\beta=0.3$, $\Delta=1$, $\alpha=1$, $\Omega=1$. Inset: the difference between the benchmark quantum--classical result [from \cref{eq:benchmark_gen}] and the quasiclassical predictions for the same methods shown in the main panel. } \label{fig:pop_SB}
\end{figure}
The long-time limits of the adiabatic population difference are shown in \cref{fig:pop_SB} as a function of the energy bias, $\varepsilon$, and are calculated with several different quasiclassical approaches. 
The other parameters are fixed to $\beta=0.3$, $\Delta=1$,  $\alpha=1$, $\Omega=1$, so that the condition for classical nuclei,  $\beta\Omega <1$, is satisfied. \cite{shi2009,zhu2013} 
This parameter choice is also justified by calculations discussed in Ref.~\onlinecite{ellipsoid}, where quasiclassical dynamics in the same parameter regime were found to agree well with numerically exact results with all degrees of freedom treated quantum mechanically. \cite{Tanimura2020HEOM,berkelbach2020} The validity of our theoretical formula, \cref{eq:long_time_gen}, for predicting the long-time limit of quasiclassical correlation functions has already been demonstrated by us for this model in another recent work, \cite{GQME} where we found that its predictions matched well with dynamical simulations. 

The general robustness of quasiclassical approaches for thermalizing correctly in any parameter regime of this model can be ascertained by considering the two parameter limits of $\varepsilon\rightarrow0$ and $\varepsilon\rightarrow\infty$. Of these extremes, one would expect the $\varepsilon\rightarrow 0$ limit to be the least challenging for quasiclassical approaches, as for our chosen parameter set of the model, the thermal energy is large compared to all other energy scales, such that the quasiclassical description of both electronic and nuclear degrees of freedom should be valid. It is therefore slightly surprising that a significant number of the approaches we tested fail in this regime, including the commonly used Ehrenfest approach. The $\varepsilon\rightarrow\infty$ limit is even more challenging, as this leads to unphysical negative populations associated with the excited state in many quasiclassical approaches [i.e., $\mathcal{C}_{\mathcal{I}z}<-1$], as can be seen in \cref{fig:pop_SB}. We do however note that the double SEO and single Wigner approaches, despite still having phase-space regions that correspond to negative populations, do thermalize to the correct physical value in the $\varepsilon\rightarrow\infty$ limit. 

The best approaches with regards to their thermalization behaviour for this model are single Wigner, SQC and MASH, which all thermalize reasonably accurately in both the $\varepsilon\rightarrow 0$ and $\varepsilon\rightarrow \infty$ limits. 
As a result, these approaches are observed to thermalize relatively well for all parameter regimes of the model, as illustrated in \cref{fig:pop_SB}.
In particular, MASH is seen to \tcb{exactly reproduce our benchmark result} across the whole range.

\subsection{An anharmonic model}\label{subsec:exp_cross}
\begin{figure}
\centering
\includegraphics[width=3.25in]{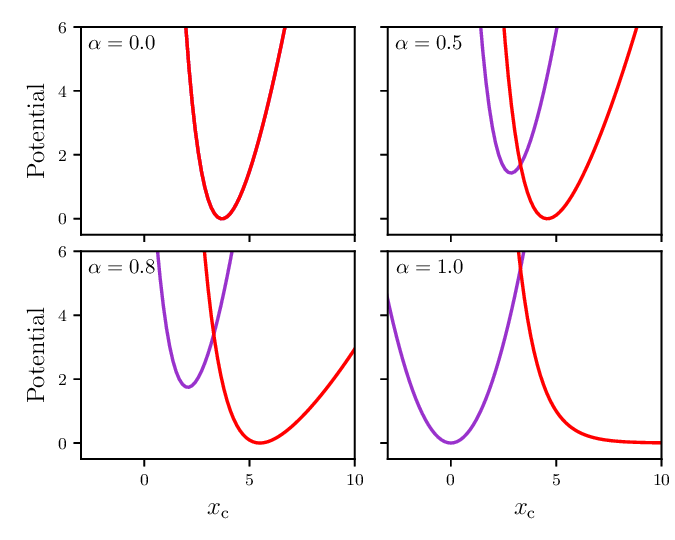}\caption{The diabatic potentials associated with the anharmonic model, $\tfrac{1}{4}[1\pm\alpha]\Omega^{2}x_{\mr c}^{2}+\tfrac{1}{2}[1\mp\alpha]\eu{-\Omega(x_{\mr c}-\bar x_{\mr{c}})}$
, for several values of the coupling parameter, $\alpha$. 
The potentials have been vertically shifted such that the lowest energy minimum of the diabats is located at zero. 
In the case of $\alpha=0$, the two diabatic potentials are identical.
}\label{fig:diss_pot}
\end{figure}

In order to fully test the effect of inverted potentials on the thermalization behaviour of mean-field approaches, an anharmonic model is needed for which at least one of the diabatic potentials can become unbounded. We choose the reaction-coordinate potentials appearing in \cref{eq:gen_pot} to be
\begin{subequations}
\label{eq:anharm_model}
\begin{align}\label{eq:pot_ep}
U_{\mathrm{RC}}(x_{\mr c})&=\tfrac{1}{2}\left[\tfrac{1}{2}\Omega^{2}x_{\mr c}^{2}+\eu{-\Omega(x_{\mr c}-\bar x_{\mathrm{c}})}\right] , \\
\hat{V}_{\mathrm{RC}}(x_{\mr c})&=\Delta\hat{\sigma}_{x}+\tfrac{\alpha}{2}\left[\tfrac{1}{2}\Omega^{2}x_{\mr c}^{2}-\eu{-\Omega(x_{\mr c}-\bar x_{\mathrm{c}})}\right]\hat{\sigma}_{z} , \label{eq:VRC_ep}
\end{align}
\end{subequations}
where $0\leq\alpha\leq1$ determines the strength of the electron-nuclear coupling and we choose the other parameters to be $\Delta=1$, $\Omega=1$, $\bar x_{\mr{c}}=5$ and $\beta=0.3$. This model is constructed so that the electronic and nuclear subsystems are uncoupled at $\alpha=0$ and the model becomes identical to a previously used 
 electronic predissociation model\cite{nonoscillatory,lawrence2018} at $\alpha=1$. On increasing $\alpha$, we see in \cref{fig:diss_pot} that the red diabatic well at lower energy becomes progressively less bounded. Additionally, the energy gap between the lower and upper adiabatic states increases,  
which more significantly drives the system into the lower energy adiabat in the long-time limit.

If a trajectory has a large enough negative population in the purple diabat [\cref{fig:diss_pot}], so that the contribution from its potential becomes inverted, then the total nuclear force can become unbounded, resulting in the trajectory accelerating off to infinity. This problem can be understood in terms of the mean-field approximation to the state-dependent potential [\cref{eq:pot_mf}], the magnitude of which can become unphysically large through the multiplication of the Cartesian mapping variables with a norm, $r$, greater than one. For this model, this results in the following definition of the effective electron--nuclear coupling associated with these methods: $\alpha_{\rm{eff}}=\alpha r$. Given that that the red diabatic potential shown in \cref{fig:diss_pot} becomes unbounded when $\alpha\geq1$, we see that this will first occur for the mean-field methods when $\alpha
 r_{\mr{max}}\geq1$, where $r_{\mr{max}}$ is the maximum allowed value of the Cartesian mapping variable norm for that method. We therefore note that the inverted potential problem is particularly problematic for MMST approaches, for which $r_{\rm{max}}=\infty$, so that the nuclear force can become unbounded for any non-zero value of $\alpha$.

\begin{figure}
\centering
\includegraphics[width=3.25in]{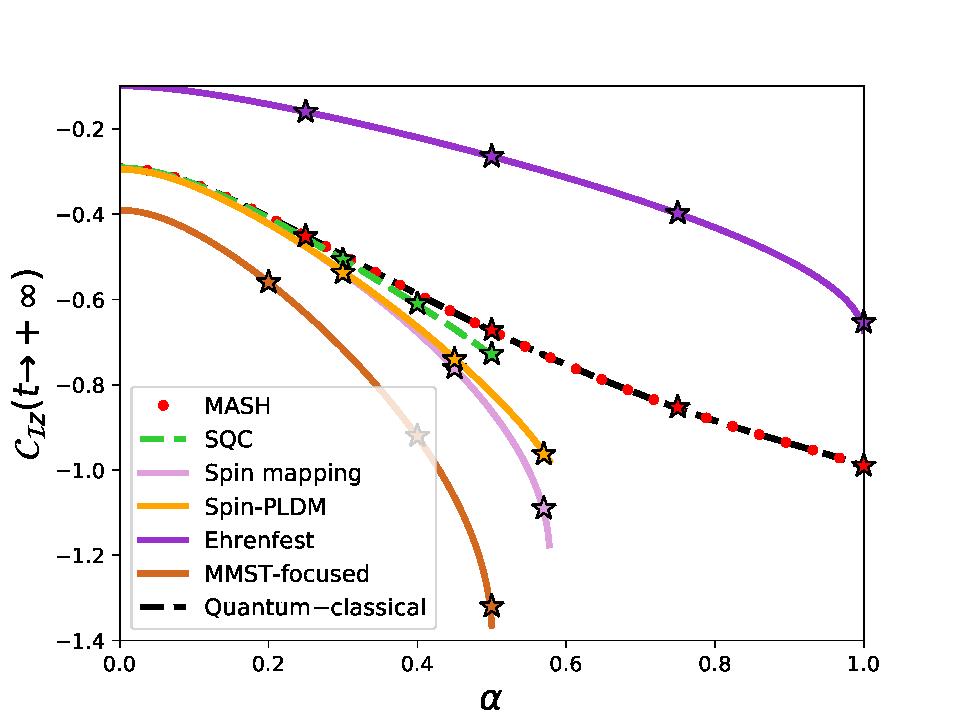}\caption{The long-time limits of the adiabatic population difference for the anharmonic model introduced in \cref{subsec:exp_cross}, as a function of the coupling parameter, $\alpha$. 
The lines in the figure correspond to our theoretical predictions given by \cref{eq:long_time_gen}, while the stars are results from dynamical simulations. 
The simulations were performed by sampling the initial nuclear phase-space variables from \cref{eq:nuc_dist} and then propagating the trajectories for $t=400$ with $\eta=2\Omega$ in \cref{eq:pure_ohm}.
Results are only shown in regions where trajectories cannot become unbounded due to inverted potentials.
}\label{fig:diss}
\end{figure}
Results for the long-time limits of the adiabatic population difference for this model are shown in \cref{fig:diss} for different quasiclassical methods.
The lines in the picture correspond to the theoretical predictions from ergodic theory, while the stars are numerical results from dynamical simulations. Results for each method are only provided for the values of $\alpha$ for which the nuclear force cannot become unbounded because of the problem of inverted potentials. We first note that the long-time limits from the dynamical simulations agree well with our simple formula [\cref{eq:long_time_gen}], further confirming the validity of applying classical ergodic theory to predict the long-time limit of quasiclassical correlation functions. Importantly, the agreement confirms that the same formula can also be used to accurately predict the long-time limit of MASH correlation functions, even though MASH does not formally have Hamiltonian generated dynamics, as discussed previously in \cref{subsec:MASH}.

Of all the quasiclassical methods we have tested, MASH is the only one in complete agreement with the quantum--classical benchmark. Its success is largely due to its windowing scheme applied consistently to both the observables and the nuclear force, such that the method does not suffer from the problem of negative populations in either. Although the SQC approach does not deviate too strongly from the exact long-time limit in the parameter regime $0\leq\alpha<0.5$ (in part due to the windowing of its observables), the problem of unbounded inverted potentials contributing to the nuclear force (which unlike for MASH are not windowed) means that SQC becomes unstable in the $0.5\leq\alpha\leq1$ regime (because $r_{\mr{max}}=2$ for SQC, as shown in \cref{app:op_SQC}). 
It would of course be possible to simply discount unbounded trajectories, but this would introduce an ad hoc modification to the method, which may affect some of its formal properties.
In addition to MASH, Ehrenfest also does not suffer from the problem of inverted potentials (because $r=1$) and can therefore also be applied in all parameter regimes of the model. However, we see from \cref{fig:diss} that the Ehrenfest long-time populations are highly inaccurate, as was also observed in \cref{fig:pop_SB}.

In \cref{sec:analysis}, we further analyze our long-time limit formula to better understand the deficiencies in the thermalization behaviour of certain methods, as well as the excellent thermalization properties of MASH.

\section{Analysis}\label{sec:analysis}
For the two-state models considered in this paper, the benchmark quantum--classical predictions for the long-time limits of the adiabatic populations  are obtained from \cref{eq:benchmark_gen} using $\hat A = \tfrac{1}{2}\hat{\mc I}_{\mr s}$ and $\hat B = \hat{\sigma}^{\text{ad}}_{z}(x)$. 
Inserting the expressions for the Hamiltonian [\cref{eq:H,eq:V}] and explicitly performing both the quantum traces and the integrals over the nuclear momenta, we find that
\begin{align}\label{eq:CIz_exact}
\mc C_{\mc Iz}^{\mathrm{QC}}(t\rightarrow\infty)=-\frac{\Braket{\sinh\left[\beta V_{z}^{\text{ad}}(x)\right]}_{\mr{b}}}{\Braket{\cosh\left[\beta V_{z}^{\text{ad}}(x)\right]}_{\mr{b}}}.
\end{align}
The phase-space average associated with the state-independent nuclear potential is defined as
\begin{equation}
\label{eq:bath_int}
\langle f(x) \rangle_{\mr{b}} = \frac{\int \mr d x\,\e^{-\beta U(x)} f(x)}{\int \mr d x\,\e^{-\beta U(x)}},
\end{equation}
and we have assumed that $\hat{\rho}_{0}(x,p)$ is normalized, $\int\rd x \rd p\,\tr_{\rm{s}}[\hat{\rho}_{0}(x,p)]=1$.
\tcb{This expression is most easily derived within the kinematic representation. 
We show in \cref{app:ham_rep} that the same result can be obtained using the Hamiltonian in the adiabatic representation, in the $\hbar\rightarrow0$ limit.}

In the following we calculate the equivalent long-time limit expressions for several quasiclassical methods. 
In order to perform the electronic phase-space integrals necessary to compare our predictions with \cref{eq:CIz_exact}, we choose to work with the spin-mapping variables in the adiabatic basis [\cref{eq:spinmap_var}]. 
\subsection{Mean-Field Approaches}\label{subsec:MF2}
For mean-field approaches, the long-time limit involves the following operator representations: 
$A(r,\tilde{\Gamma})=\tfrac{1}{2}\mathcal{I}_{\mr{s}}(r)$, $B(r,\tilde{\Gamma})=rS^{\mr{ad}}_{z}$ and $\rho_{0}(r,\tilde{\Gamma})=\rho_{0,\mr{s}}(r)\rho_{\mr{b}}(x,p)$.
Inserting these into \cref{eq:long_time_gen} and performing some of the phase-space integrals, we find
\begin{subequations}
\label{eq:ad_lim_MF2}
\begin{align}
&\mc C^{\text{MF}}_{\mc I z}(t\rightarrow\infty)=-\int_{0}^{\infty}\rd r \, r^{2}\rho_{0,\text{s}}(r)\,\mathcal{I}_{\text{s}}(r) \nonumber \\
&\times\Bigg\langle\frac{\beta rV^{\text{ad}}_{z}(x)\cosh[\beta rV^{\text{ad}}_{z}(x)]-\sinh[\beta rV^{\text{ad}}_{z}(x)]}{[\beta rV_{z}^{\text{ad}}(x)]^2\mathcal{Z}^{\mr{MF}}(r)}\Bigg\rangle_{\mr{b}}
, \\
&\mathcal{Z}^{\mr{MF}}(r)=\Braket{\frac{\sinh{\left[\beta rV_{z}^{\text{ad}}(x)\right]}}{\beta r V^{\text{ad}}_{z}(x)}}_{\mr{b}}. \label{eq:Z_MF}
\end{align}
\end{subequations}

Comparing \cref{eq:CIz_exact} and \cref{eq:ad_lim_MF2}, we see that the expression for the long-time limit of the mean-field correlation function has the wrong functional form, such that there is no universal expression for $\rho_{0,\text{s}}(r)\mathcal{I}_{\text{s}}(r)$ that guarantees that the corresponding mean-field approach would always thermalize to the \tcb{quantum--classical benchmark}. 
For a given temperature, a system-specific expression for these functions can however be determined so that \cref{eq:CIz_exact} and \cref{eq:ad_lim_MF2} agree. This idea has recently been used to design the ``ellipsoid mapping'', a new mean-field approach for computing thermal correlation functions that rigorously obeys detailed balance. \cite{ellipsoid} Such an approach however requires a static calculation to be carried out at thermal equilibrium for each system and temperature of interest, in order to calculate the correct associated expression for $\rho_{0,\text{s}}(r)\mathcal{I}_{\text{s}}(r)$, before any dynamical simulations can be performed.

If we instead decide to only use universal functions for $\rho_{0,\text{s}}(r)\mathcal{I}_{\text{s}}(r)$, then their form can at best be chosen to ensure correct thermalization in certain parameter limits. In the following, we consider the limiting cases of high temperature, low temperature and weak system--bath coupling. For the high-temperature limit ($\beta\rightarrow0$), we show in \cref{subsec:beta_to_zero} that mean-field approaches are at best capable of reproducing the \tcb{benchmark} [\cref{eq:CIz_exact}] up to first order in $\beta$.
To achieve this for any two-level system, the condition
\begin{equation}\label{eq:get1order}
\frac{1}{3}\int_{0}^{\infty}\rd r\,r^{3}\rho_{0,\text{s}}(r)\mathcal{I}_{\text{s}}(r)=1 
\end{equation}
must be satisfied.
Mean-field methods that fulfill \cref{eq:get1order} are marked with a tick in the $\beta\rightarrow0$ column of \cref{tab:comparison}. For methods that do not satisfy this condition, the value associated with the left-hand side of \cref{eq:get1order} is given, which by comparing \cref{eq:ad_lim_ex2} and \cref{eq:MF_exp2} is seen to be the multiplicative error in the long-time population difference of the method when $\beta\rightarrow0$. 
Incidentally, these errors qualitatively explain the deviations of the long-time limit populations from the quantum--classical benchmark in the $\varepsilon\rightarrow0$ limit of \cref{fig:pop_SB}. While the $\varepsilon\rightarrow0$ and the $\beta\rightarrow0$ limits of a spin--boson model do not always coincide, they do so approximately for the parameter regime that we study here.

In the low-temperature limit ($\beta\to\infty$), mean-field approaches can correctly reproduce the long-time limit of correlation functions to at best zeroth order in $\eu{-\beta V_{z}^{\mr{ad}}(x)}$. As shown in \cref{subsec:beta_to_inf}, they must then satisfy
\begin{equation}\label{eq:get1order_cold}
\int_{0}^{\infty}\rd r\, r^{2}\rho_{0,\text{s}}(r)\mathcal{I}_{\text{s}}(r)=1 .
\end{equation}
Mean-field methods that satisfy \cref{eq:get1order_cold} are marked with a tick in the $\beta\rightarrow\infty$ column of \cref{tab:comparison}. For methods that do not satisfy this condition, the value associated with the left-hand side of \cref{eq:get1order_cold} is given, which is the multiplicative error in the long-time population difference of the method when $\beta\rightarrow\infty$.
For strongly asymmetric systems, relaxation in the long-time limit will predominately occur into the lowest potential-energy well, like in the low-temperature limit. This means that our results for $\beta\rightarrow\infty$ can also be used to explain the thermalization behaviour of mean-field methods in the spin--boson model for $\varepsilon\rightarrow\infty$, given in \cref{fig:pop_SB}. We note however that the asymptotic approach to the low-temperature limit, which corresponds to the term that is first-order in $\eu{-\beta V_{z}^{\mr{ad}}(x)}$, is always wrong for mean-field approaches, as is also observed numerically from our results in \cref{fig:pop_SB}. 

We show in \cref{app:alpha_to_zero} (and also illustrate with crosses for all the mean-field methods in the $\alpha\rightarrow0$ column of \cref{tab:comparison}) that it is not possible to design mean-field methods which consistently thermalize correctly in the weak-coupling limit. Despite this, spin mapping and spin-PLDM appear very accurate in the $\alpha\rightarrow0$ limit of the anharmonic model [see \cref{fig:diss}]. This is because  the weak-coupling limit for this particular parameter set of the model also coincides with the $\beta\rightarrow0$ limit.
These methods would not have been as accurate in this limit if we would have instead studied a lower-temperature regime. 
Let us remark that if the electron--nuclear coupling is identically zero, all methods discussed are able to correctly capture the Rabi oscillations of the isolated electronic subsystem.
In this case ergodic theory does not apply, given that the dynamics do not thermalize to equilibrium at long times.
Note that this condition is different from the $\alpha\to 0$ limit discussed in \cref{tab:comparison,app:alpha_to_zero}, where the coupling is assumed to be infinitesimal but nonzero.

\subsection{Symmetric quasiclassical windowing (SQC)}\label{subsec:SQC2}
For SQC, the potential still retains the same representation as for the mean-field methods [i.e., $\hat{\sigma}^{\mr{ad}}_{z}(x)\mapsto rS^{\mr{ad}}_{z}$ in \cref{eq:V}], but the observable operators are now represented by the triangular windows, given by \cref{eq:SQCop} in terms of the spin-mapping variables. Inserting these operator representations into \cref{eq:long_time_gen} and performing some of the phase-space integrals gives
\begin{subequations}\label{eq:limSQC}
\begin{align}
\mc C^{\mr{SQC}}_{\mathcal{I}z}(t\rightarrow\infty)&=\frac{\tilde{\mc C}^{\mr{SQC}}_{\mathcal{I}z}(t\rightarrow\infty)}{\mathcal{N}^{\mr{SQC}}(t\rightarrow\infty)}, \label{eq:ratioSQC}\\
\tilde{\mc C}^{\mr{SQC}}_{\mathcal{I}z}(t\rightarrow\infty)&=-\int_{1}^{2}\rd r\,(r-1) \nn\\
&\times\Braket{\frac{\sinh[\beta V^{\mathrm{ad}}_{z}(x)]\sinh[\beta (r-1)V^{\mathrm{ad}}_{z}(x)]}{\beta rV^{\mathrm{ad}}_{z}(x)\mathcal{Z}^{\mr{MF}}(r)}}_{\mr{b}},\label{eq:tCIz_SQC}
 \\
\mathcal{N}^{\mr{SQC}}(t\rightarrow\infty)&=\int_{1}^{2}\rd r\,(r-1)  \nn\\
&\times\Braket{\frac{\cosh[\beta V^{\mathrm{ad}}_{z}(x)]\sinh[\beta (r-1)V^{\mathrm{ad}}_{z}(x)]}{\beta r V^{\mathrm{ad}}_{z}(x)\mathcal{Z}^{\mr{MF}}(r)}}_{\mr{b}},\label{eq:N_SQC}
\end{align}
\end{subequations}
where $\mathcal{N}^{\mr{SQC}}(t)$ is the normalization factor, which is required to ensure that the electronic populations sum to one.
Note that we have applied the ergodic theory outlined in \cref{subsec:erg} separately to each term in the ratio of \cref{eq:ratioSQC}.

A comparison with \cref{eq:CIz_exact} clearly shows that \cref{eq:limSQC} is not exact, again having the wrong functional form compared to the \tcb{benchmark} quantum--classical long-time limit. 
However, the advantage of SQC over the other mean-field approaches is that the long-time limit of its correlation function also \tcb{reproduces the benchmark} in the weak-coupling limit,\cite{Miller2015SQC}
in addition to the high- and low-temperature limits (as indicated in \cref{tab:comparison})\@. More details on the thermalization of SQC in these limits is given in \cref{app:limits_lt}\@. The fact that SQC can simultaneously describe the thermalization correctly in all of these limits explains why it is reasonably accurate for all values of $\varepsilon$ for the spin--boson model in \cref{fig:pop_SB} 
and accurate in the $\alpha\rightarrow0$ limit of the anharmonic model in \cref{fig:diss_pot}.  
However for larger values of $\alpha$, there is a clear deviation even before the unbounded inverted potentials appear at $\alpha\ge0.5$.

\subsection{Mapping approach to surface hopping (MASH)}\label{subsec:MASH2}
\begin{figure}
\centering
\includegraphics[width=3.25in]{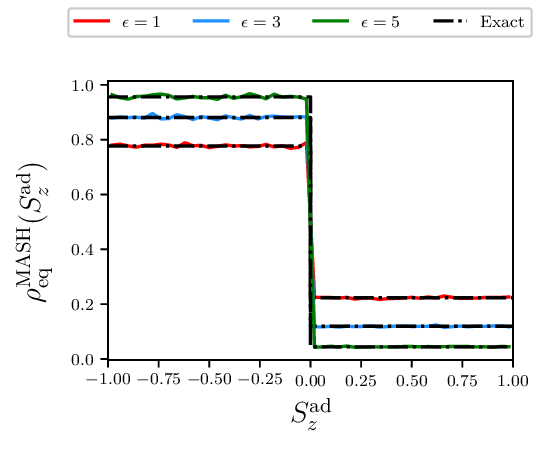}
\caption{The marginal equilibrium distribution function for the mapping variable $S_{z}^{\text{ad}}$, calculated for the spin--boson model defined in \cref{subsec:sys_bath}, for several values of the energy bias $\varepsilon$. 
The colored lines are obtained from histogramming MASH trajectories initialized from \cref{eq:nuc_dist} and propagated for $t=500$ with the optimal damping parameter, 
$\eta=2\Omega$ [\cref{eq:pure_ohm}]. Additionally, the black dashed lines correspond to the expected results from classical ergodic theory, given by \cref{eq:MASH_dist}.}\label{fig:dist_MASH}
\end{figure}

\begin{subequations}\label{eq:MASH}
In MASH, both the potential and observable operators are consistently described through the windowing scheme, $\hat{\sigma}^{\mr{ad}}_{z}(x)\mapsto\sgn(S_{z}^{\mr{ad}})$.
Following a strategy similar to the one outlined above for the other methods, we find for the long-time population difference of MASH that
\begin{align} \label{eq:MASH_ltl}
\mc C^{\text{MASH}}_{\mc I z}(t\rightarrow\infty)&=\int_{-1}^{1}\rd S_{z}^{\text{ad}}\,\sgn{(S_{z}^{\text{ad}})}\rho^{\mathrm{MASH}}_{\mathrm{eq}}(S_{z}^{\mathrm{ad}}) , \\
\rho^{\mathrm{MASH}}_{\mathrm{eq}}(S_{z}^{\mathrm{ad}})&=\frac{\braket{\eu{-\beta V_{z}^{\mathrm{ad}}(x)\sgn(S^{\mathrm{ad}}_{z})}}_{\mr{b}}}{2\Braket{\cosh{[\beta V_{z}^{\text{ad}}(x)]}}_{\mr{b}}} , \label{eq:MASH_dist}
\end{align}
where we have additionally used that the initial distribution for the spin-mapping variables for this correlation function is $r\rho_{0,\mr{s}}=\mathcal{W}_{\mr{PP}}(\bm{S}^{\mr{ad}})=2|S_{z}^{\mr{ad}}|$.\cite{MASH} The marginal equilibrium distribution function for MASH, $\rho_{\mr{eq}}^{\mr{MASH}}(S_{z}^{\mr{ad}})$, corresponds to the long-time limit of the conditional probability, given by \cref{eq:mixing}, with the bath degrees of freedom integrated out.

We first note that our expression for the marginal equilibrium distribution function for MASH agrees with the long-time distribution of $S^{\mr{ad}}_{z}$, histogrammed from an ensemble of dynamical MASH trajectories. This is shown in \cref{fig:dist_MASH} for the same spin--boson model that was introduced in \cref{subsec:sys_bath}, for several values of $\varepsilon$. This marginal distribution constitutes an exact discrete-valued representation of the quantum--classical Boltzmann factors
, which demonstrates how MASH effectively quantizes the electronic states, enabling it to thermalize to the \tcb{benchmark} quantum--classical distribution in all cases. 
The ultimate proof of the long-time accuracy of the MASH dynamics is the fact that \cref{eq:CIz_exact} and \cref{eq:MASH_ltl} are seen to be identical once the integrals over the spin-mapping variables have been performed analytically.
To the best of our knowledge, MASH is the only nonequilibrium quasiclassical approach that is guaranteed  to thermalize correctly in all nonadiabatic systems \tcb{in the $\hbar\rightarrow0$ limit}.

\end{subequations}

This result may come as a surprise as it is known that MASH does not rigorously obey detailed balance, due to the non-conservation of its weighting factor by the dynamics.\cite{MASH} This raises the question of how MASH can still thermalize correctly. 
To describe the issue quantitatively, a measure of the microscopic reversibility error (MRE) was defined in Ref.~\onlinecite{MASH}. 
Based on the same ergodicity arguments used here, we show in \cref{app:microMASH} that in the long-time limit the MRE vanishes.
This shows that the violation of detailed balance by MASH can only lead to errors at intermediate times, with no adverse effects on its long-time relaxation behaviour.


\section{Conclusions}
In this paper, we presented an analysis of the long-time limits of quasiclassical approaches for simulating nonadiabatic dynamics.
We exploited the Hamiltonian structure of the dynamics to make use of results from classical ergodic theory, rigorously accounting for the conservation of the norm of the mapping variables.
This allowed us to assess and compare the accuracy of different methods in capturing the correct detailed balance at long times. 
Our theoretical framework can be applied to any quasiclassical method which is incompressible (i.e., it conserves phase-space volume) non-integrable (i.e., the dynamics cannot be solved analytically)
and deterministic.\cite{arnold1978} 

Using our theory, we tested the thermalization behaviour of a wide array of quasiclassical approaches on both the harmonic and anharmonic models.
Our analysis revealed that most of the commonly used quasiclassical approaches violate detailed balance and thus do not recover the correct quantum--classical thermal distribution. Errors are particularly large for the anharmonic model, where the problem of inverted potentials meant that trajectories from the majority of quasiclassical approaches were unphysically accelerated off to infinity. 
Among all of the methods considered, only MASH is guaranteed to predict the exact quantum--classical correlation functions in the long-time limit, \tcb{to zeroth-order in $\hbar$}.
This method is therefore capable of solving the long-standing issue of detailed balance (at least in the long-time limit), a problem which has plagued the field of quasiclassical dynamics to date. In addition, because MASH does not suffer from the problem of inverted potentials, it is ideally suited for performing ab-initio nonadiabatic simulations of molecules, where the strongly anharmonic adiabatic potentials would pose a significant challenge to the other quasiclassical approaches.  

The main limitation of the original version of MASH, introduced in Ref.~\onlinecite{MASH}, is that it was derived only for two-state problems. 
Recently a multi-state generalization of MASH has been proposed by Runeson and Manolopoulos.\cite{runeson2023}
In this version, they modified the observable operators to be more similar to the mean-field approaches, which made it easier for them to generalize the underlying theory to multi-state problems.
Like the original MASH approach, it is guaranteed to thermalize correctly in the long-time limit. 
However, it is not guaranteed to exactly reproduce the short-time dynamics of the quantum--classical Liouville equation, which was the means by which the original MASH approach obtained unique and rigorous prescriptions for the momentum rescalings, treatment of frustrated hops and decoherence corrections. In addition, the measurement of observables is not performed using the same windowing procedure as for the nuclear force, meaning that the two can become inconsistent and the observables may measure negative populations. Whilst the accuracy of the numerical results look promising,\cite{runeson2023} given that the rigorous nature and internal consistency of the original MASH approach was its main advantage over FSSH, we still think that there is more work to be done in order to obtain the ultimate multi-state generalization of MASH. 


There may however still be a place for mean-field mapping methods.
In contrast to typical molecules in the gas phase, 
many solid-state systems are characterized by a high degree of electronic coherence.
Due to the dense bands of electronic states, the treatment of the nonadiabatic dynamics using mean-field approaches seems the most suitable. From our analysis, we conclude that single-Wigner and SQC would be the best mean-field approaches to use in this case. Given that the electronic structure of solids is fairly harmonic, we also expect that the problem of inverted potentials would not be as severe.

It is interesting to compare the conclusions from our analysis to those obtained in previous work. In Refs.~\onlinecite{identity,FMO,linearized}, it was observed that the unity methods led to improved accuracy compared to other MMST approaches. Similar improvements were observed in Refs.~\onlinecite{spinmap,multispin,spinPLDM1,spinPLDM2,nonlinear} for spin-mapping methods, which also treat the identity operator exactly within the mapping. However, most previous studies were performed on harmonic models with reasonably small energy biases. From our analysis in this paper, we observe that for more extreme systems, the quasiclassical approaches that treat the identity operator exactly do not always perform the best, suggesting that other criteria may also be important to consider. For example, a recent analysis performed on the ab-initio exited-state dynamics in ethylene found that the best approaches in this case were those that had the least phase-space volume associated with negative populations.\cite{miyazaki2023}

The long-time accuracy of quasiclassical methods can often be significantly improved by utilizing these techniques within the Nakajima--Zwanzig generalized quantum master equation (GQME) formalism. \cite{Kelly2016master,Mulvihill2019LSCGQME,Montoya2016GQME}
This approach describes the non-unitary dissipative dynamics of the subsystem in terms of a non-Markovian equation of motion.
Memory effects in the GQME are captured by kernels that decay on timescales that are in general much shorter than the typical electronic relaxation times. 
This means that it is often advantageous to calculate the kernels using quasiclassical methods, instead of obtaining the full dynamics directly. However it has recently been shown that using the GQME is not guaranteed to fix detailed balance\cite{GQME} if the input quasiclassical method is not sufficiently accurate at short times.
Given the remarkable long-time accuracy exhibited by MASH, it could be insightful to couple this method to the GQME for determining the importance of non-Markovian dissipative effects in the dynamics, as well as to reduce the cost of quasiclassical simulations through computing short-lived memory kernels.

Recently there has been increased interest in understanding the effect of strong light-matter coupling on the properties of matter. For quantum light, the coupled electron--photon dynamics can be described using quasiclassical approaches, where the photon phase-space variables are initially sampled from a Wigner distribution.\cite{Hoffmann2019,Hoffmann2019polariton,Saller2021polariton} While the short-time dynamics is observed to be fairly accurate, the long-time dynamics is known to suffer as a result of unphysical zero-point energy leakage from these quantum modes.\cite{Hsieh2023}
For classical light, the presence of an external driving field breaks the thermalization to the Boltzmann distribution. The dynamics converge at long times to nonequilibrium stationary states which can be described in certain parameter regimes by the Floquet--Gibbs distribution. \cite{engelhardt2019,shirai2015}
In both cases, our analysis based on classical ergodic theory could be extended in order to analyze how well quasiclassical approaches describe the correct light-modified long-time properties.


\section*{Acknowledgments}
This project has received funding from European Union’s Horizon $2020$ under MCSA Grant No.~$801459$, FP-RESOMUS\@. We also thank Johan Runeson and Joseph Lawrence for helpful discussions. 

\section*{Author Declarations}
\subsection*{Conflict of Interest}
The authors have no conflicts to disclose.
\subsection*{Author Contributions}
\textbf{Graziano Amati}: Conceptualization (equal); Formal analysis (equal); Investigation (equal); Methodology (equal); Visualization (equal); Writing -- original draft (equal); Writing -- review \& editing (equal).
\textbf{Jonathan R. Mannouch}: Conceptualization (equal); Formal analysis (equal); Investigation (equal); Methodology (equal); Visualization (equal); Writing -- original draft (equal); Writing -- review \& editing (equal).
\textbf{Jeremy O. Richardson}: Conceptualization (equal); Formal analysis (supporting); Supervision (lead); Writing -- review \& editing (equal).


\begin{appendix}
\tcb{\section{Hamiltonian representations}\label{app:ham_rep}
The total Hamiltonian can be expressed in various representations. 
In the following we demonstrate that our results are independent of these choices.
}

\tcb{For simplicity, we perform the analysis in this section on two-state Hamiltonians, although our conclusions are entirely valid for multi-state problems. We start with the Hamiltonian written in the diabatic basis [\cref{eq:H}]. 
If the classical nuclear limit is first taken before the electronic operators are transformed to the adiabatic basis, we end up in the kinematic representation\cite{Cotton2017mapping} [\cref{eq:V}]
\begin{equation}
\label{eq:H_kin}
\hat{H}_{\text{kin}}(x,p)=\frac{p^{2}}{2m}+U(x)+V^{\rm{ad}}_{z}(x)\hat{\sigma}^{\rm{ad}}_{z}(x) .
\end{equation}}

\tcb{To illustrate that our quantum--classical benchmark [\cref{eq:benchmark_gen}] is identical in both the diabatic and kinematic representations, we compute the quantum--classical partition function using the diabatic representation 
\begin{equation}
\label{eq:partition_diab}
\begin{split}
    Z_\mathrm{QC} &= \int \rd x \rd p \,\tr_\mathrm{s}[\eu{-\beta \hat{H}(x,p)}] 
    \\ &= \int \rd x \rd p \, \eu{-\beta p^2/2m-\beta U(x)} \tr_\mathrm{s}\left[\eu{-\beta \hat{V}(x)}\right] 
    \\ &= \int \rd x \rd p \, \eu{-\beta p^2/2m-\beta U(x)}\left(\eu{-\beta V^{\mr{ad}}_{z}(x)}+\eu{\beta V^{\mr{ad}}_{z}(x)}\right) ,
\end{split}
\end{equation}
where we have used the fact that the trace of the exponential of an operator is the sum of the exponentials of its eigenvalues.
This result is clearly identical to that obtained from the kinematic representation, as the only difference is that the potential energy is already diagonal in the latter.}


\tcb{
If the order of operations is interchanged so that the transformation to the adiabatic basis is done before the classical nuclear limit is taken, we end up in the adiabatic representation
\begin{equation}
\label{eq:H_ad}
\begin{split}
\hat{H}_{\text{ad}}(x,p_{\mr{ad}})=\frac{1}{2m}&\left(p_{\mr{ad}}+\hbar d(x)\hat{\sigma}^{\rm{ad}}_{y}(x)\right)^{2} \\
&+U(x)+ V^{\rm{ad}}_{z}(x)\hat{\sigma}^{\rm{ad}}_{z}(x) ,
\end{split}
\end{equation}
where $d(x)=\Braket{\psi_{+}(x)|\frac{\partial}{\partial x}|\psi_{-}(x)}$ is the nonadiabatic coupling vector and $p_{\mr{ad}}$ is the canonical momentum associated with the adiabatic representation. Factors of $\hbar$, which originate from the nuclear momentum operator, are explicitly retained for the following discussion. Computing the quantum--classical partition function using the adiabatic representation, we find
\begin{equation}
\label{eq:partition_ad}
\begin{split}
   Z_\mathrm{QC}^\mathrm{ad} &= \int \rd x \rd p_\mathrm{ad}\, \tr_\mathrm{s}\left[\eu{-\beta \hat{H}_\mathrm{ad}(x,p_\mathrm{ad})}\right] ,
    \\ &= \int \rd x \rd p_\mathrm{ad} \, \eu{-\beta (p_\mathrm{ad}^2+\hbar^2d(x)^2)/2m-\beta U(x)} 
    \\&\quad\times\tr_\mathrm{s}\left[\eu{-\beta\left(\hbar d(x)p_\mathrm{ad}\hat{\sigma}_{y}^{\mr{ad}}(x)/m + \hat{V}_z^{\mr{ad}}(x)\hat{\sigma}_z^{\mr{ad}}(x)\right)}\right] ,
    \\&\simeq \int \rd x \rd p_\mr{ad} \, \eu{-\beta p_\mr{ad}^2/2m-\beta U(x)} \tr_\mr{s}\left[\eu{-\beta V_z^\mr{ad}(x)\hat\sigma_z^\mr{ad}(x)}\right]
\end{split}
\end{equation}
where in the final line we take the $\hbar\rightarrow0$ limit. Is it then seen that \cref{eq:partition_diab,eq:partition_ad} are identical in this limit.
}

\section{Operators for SQC in spherical coordinates}\label{app:op_SQC}
In this appendix, we rewrite the two-state SQC representation of the adiabatic population operators in spherical coordinates. Using this result, an analytic expression for the long-time limit of the adiabatic populations can be obtained for this method [\cref{eq:limSQC}]. 
Unlike most other mapping approaches, the exact form of SQC depends on the representation chosen.
We will utilize the adiabatic representation, which has previously been employed in ab initio simulations,\cite{Talbot2022SQCDFT}
because this is expected to lead to the best possible predictions available from the approach.
It is then also easier to compare with MASH, which is always run in the adiabatic representation.

The SQC approach is usually expressed in terms of action--angle variables. These can be transformed to spherical coordinates by combining \cref{eq:AA,eq:spinmap_var}, which leads to
\begin{subequations}\label{eq:aa_to_sph}
\begin{align}
rS^{\mr{ad}}_{x}&=r\sin\theta\cos\phi =2\sqrt{(n_{+}+\gamma)(n_{-}+\gamma)}\cos(q_+-q_-) , \\ rS^{\mr{ad}}_{y}&=r\sin\theta\sin\phi = 2\sqrt{(n_{+}+\gamma)(n_{-}+\gamma)}\sin(q_+-q_-) , \\
rS^{\mr{ad}}_{z}&=r\cos\theta = n_{+}-n_{-} , 
\end{align}
\end{subequations}
where 
$q_+$ and $q_-$ are
the angle variables for the upper and lower adiabatic states and are initially sampled uniformly in the range $[0,2\pi)$. 
Additionally, $n_{+}$ and $n_{-}$ are the action variables for the two adiabatic states, sampled from the appropriate triangular window corresponding to the initial electronic population [\cref{eq:window}].\cite{Cotton2016SQC}

\begin{figure}
		\centering
		\includegraphics[width=3.25in]{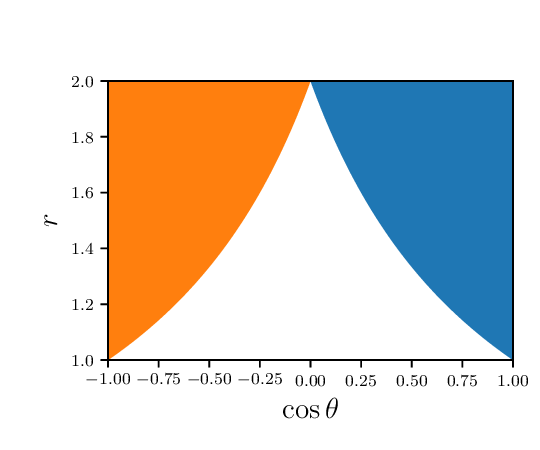}\caption{The triangular windows used in SQC expressed in spherical polar coordinates. The blue and orange regions correspond to $W_{+}(r,\theta)$ and $W_{-}(r,\theta)$ respectively, which take the form of spherical caps. }\label{fig:window_spherical}
\end{figure}

The expression for the windows in spherical coordinates can be determined from the inverse transformation of \cref{eq:aa_to_sph}, which corresponds to
\begin{subequations}\label{eq:sph_to_aa}
\begin{align}
n_{+}&=\frac{r}{2}(1 + \cos\theta)-\gamma , \\
n_{-}&=\frac{r}{2}(1 - \cos\theta)-\gamma, \\
q_+-q_-&=\phi .
\end{align}
\end{subequations}
Note that $q_++q_-$ is an unimportant cyclic variable conjugate to the conserved radius.
The new angle variable $\phi$ is initially sampled uniformally in the range $[0,2\pi)$. 
Finally, inserting \cref{eq:sph_to_aa} into \cref{eq:window}, we find that the triangular windows in spherical coordinates are
\begin{subequations}
\label{eq:window_spherical}
\begin{align}
\mc W_+(r,\theta) &= h\left(\tfrac{r}{2}(1+\cos\theta)-1 \right)h(2-r),\\
\mc W_-(r,\theta)&= h\left(\tfrac{r}{2}(1-\cos\theta)-1\right)h(2-r).
\end{align}
\end{subequations}
which are illustrated in \cref{fig:window_spherical}. Note that for $r=2$, the SQC windows fill the $\cos\theta$ axis and correspond to the hemispheres of the spin-sphere, which are identical to the windows used in MASH\@. This connection is unsurprising, as it has already been noted that the weighting factors used in the MASH correlation functions correspond to those for the $r=2$ MMST sphere.\cite{MASH} Finally, in order to find the SQC representations for the $r^{N-1}\rho_{0,\mr{s}}(r)$ distribution and the $\hat{\mathcal{I}}_{\mr{s}}$ and $\hat{\sigma}_{z}^{\mr{ad}}(x)$ operators, linear combinations of \cref{eq:window_spherical} can be taken, leading to 
\begin{subequations}\label{eq:SQCop}
\begin{align}
r^{N-1}\rho_{0,\mr{s}}(r)&= h(2-r) , \label{eq:SQC_indent} \\
\mathcal{I}_{\mr s}(r)&= h(|rS^{\text{ad}}_{z}|-2+r) , \label{eq:SQC_indent2} \\
\sigma^{\text{ad}}_{z}(x)&=\sgn(S_{z}^{\text{ad}})h(|rS^{\text{ad}}_{z}|-2+r) .
\end{align}
\end{subequations}

\section{Perturbative expansions of the long-time limit}\label{app:limits_lt}
In this appendix, we obtain perturbative expressions for the long-time limit of various quasiclassical correlation functions.
These expressions allow us to assess how accurately different methods thermalize in different parameter regimes. 
Note that it is unnecessary to carry out the following analysis for MASH, because, as proven in \cref{subsec:MASH2}, the long-time limits of its correlation functions are exact and hence all orders of these perturbative expansions are guaranteed to be correctly reproduced.
\subsection{The $\beta\rightarrow 0$ limit}\label{subsec:beta_to_zero}
The $\beta\rightarrow 0$ limit allows us to assess how accurately quasiclassical approaches reproduce the long-time limit of correlation functions at high-temperature.
Because we are already assuming in this paper that we are in the high-temperature limit with respect to the nuclear degrees of freedom (so that a classical treatment of the bath is valid), we clarify that by the high-temperature limit here we mean relative to the electronic energy scales. 

Using the expressions \cref{eq:CIz_exact,eq:ad_lim_MF2,eq:limSQC}, we obtain the following results for the first-order term in the corresponding high-temperature Taylor expansions
\begin{subequations}
\label{eq:beta_to_0}
\begin{align}
\mathcal{C}_{\mathcal{I}z}^{\mr{QC}}(t\rightarrow\infty) &\sim - \beta\langle V_z^{\mr{ad}}(x) \rangle_{\mr b} \label{eq:ad_lim_ex2} , \\
 \mc C^{\text{MF}}_{\mathcal{I}z}(t\rightarrow\infty)&\sim-\frac{\beta}{3}\langle   V^{\mathrm{ad}}_z(x) \rangle_{\mr b}\int\rd r\, r^{3}\rho_{0,\text{s}}(r)\,\mathcal{I}_{\text{s}}(r) \label{eq:MF_exp2} ,\\
 C^{\text{SQC}}_{\mathcal{I}z}(t\rightarrow\infty)&\sim- \beta\langle V_z^{\mr{ad}}(x) \rangle_{\mr b} . \label{eq:ad_lim_sqc2}
\end{align}
\end{subequations}
We thus see that SQC is able to correctly thermalize in the high-temperature limit, although we note that this is not the case for all mean-field methods. Comparing \cref{eq:ad_lim_ex2,eq:MF_exp2} leads to the condition expressed in \cref{eq:get1order} that must be satisfied for mean-field methods to describe this limit correctly.


While the $\varepsilon\rightarrow0$ and the $\beta\rightarrow0$ limits of a spin--boson model do not always coincide, they approximately do for the parameter regime that we study here. This means that these high-temperature limit formulas can be used to understand the thermalization behaviour of quasiclassical approaches in the $\varepsilon\rightarrow 0$ limit of the spin-boson model discussed in \cref{subsec:sys_bath}.
\subsection{The $\beta\rightarrow\infty$ limit}\label{subsec:beta_to_inf}
Here we investigate the low-temperature ($\beta\rightarrow\infty$) limit with respect to the electronic energy scales. We still assume that the temperature is high with respect to the nuclear degrees of freedom, such that it is valid to treat them classically. The zeroth-order terms of these expansions are
\begin{subequations}
\label{eq:beta_to_inf}
\begin{align}
 \mathcal{C}_{\mathcal{I}z}^{\mr{QC}}(t\rightarrow\infty)&\sim-1 ,\label{eq:CIz_lowTex} \\
 \mc C^{\text{MF}}_{\mathcal{I}z}(t\rightarrow\infty)&\sim-\int\rd r \,r^{2}\rho_{0,\text{s}}(r)\,\mathcal{I}_{\text{s}}(r),\label{eq:CIz_lowTMF} \\
\mathcal{C}_{\mathcal{I}z}^{\mr{SQC}}(t\rightarrow\infty)&\sim-1 .
\end{align}
\end{subequations}
We note that SQC also describes the thermalization in the low-temperature regime correctly, while mean-field methods do not unless they satisfy the condition \cref{eq:get1order_cold}. 

In the long-time limit, the system will predominately relax into the lowest potential-energy well in both the $\beta\rightarrow\infty$ and the $\varepsilon\rightarrow\infty$ limits of the spin--boson model. This means that our long-time limit formulas for ${\beta\rightarrow\infty}$ can also be used to understand the thermalization behaviour of quasiclassical approaches in the $\varepsilon\rightarrow\infty$ limit of the spin--boson model discussed in Section~\ref{subsec:sys_bath}. 
We additionally remark that quasiclassical approaches that reproduce the correct $\beta\rightarrow\infty$ limit are not in general guaranteed to reproduce the right asymptotic approach (apart from MASH), as observed in our numerical results [see in particular Ehrenfest and double SEO in \cref{fig:pop_SB}].
\subsection{The $\alpha\rightarrow 0$ limit}\label{app:alpha_to_zero}
The weak-coupling limit ($\alpha\rightarrow0$) leads to a significant simplification of our long-time limit formulas, because in this limit $V^{\mr{ad}}_{z}(x)=V_{z}^{(0)}$ becomes independent of the nuclear positions and such terms can be taken outside the phase-space averages, $\braket{\cdots}_{\mr{b}}$. This means that identical terms in the numerator and denominator of a fraction that previously belonged in different phase-space integrals now cancel. Performing these simplifications leads to the zeroth-order terms
\begin{subequations}
\label{eq:lambda_to_0}
\begin{align}
\mathcal{C}_{\mathcal{I}z}^{\mr{QC}}(t\rightarrow\infty) &\sim -\tanh{(\beta V_{z}^{(0)}})
 , \\
 \mc C^{\text{MF}}_{\mathcal{I}z}(t\rightarrow\infty)&\sim-\int\rd r\, r^{2}\rho_{0,\text{s}}(r)\,\mathcal{I}_{\text{s}}(r)\nn \\
 &\qquad \times\Bigg[\coth(\beta r V_{z}^{(0)})-\frac{1}{\beta r V_{z}^{(0)}}\Bigg] 
 ,\label{eq:CMF_small_alpha}\\
 C^{\text{SQC}}_{\mathcal{I}z}(t\rightarrow\infty)&\sim-\tanh{(\beta V_{z}^{(0)}}) .
\end{align}
\end{subequations}
We find that SQC is also exact in the weak-coupling limit, as was also shown in Ref.~\onlinecite{Miller2015SQC}.\footnote{Our derivation is similar to that used in Ref.~\onlinecite{Miller2015SQC}, but we believe it to be more rigorous as we explicitly take account of the conserved norm of the mapping variables.}
This time, the mean-field condition depends on $\beta V_{z}^{(0)}$ and so cannot be satisfied in general unless the method is explicitly dependent on this system-dependent variable
(similar to Ref.~\onlinecite{ellipsoid}). 
These results are summarized in the corresponding column of \cref{tab:comparison}.

\section{Microscopic-reversibility error in MASH}\label{app:microMASH}
The \emph{microscopic-reversibility error} (MRE) was introduced as a way of estimating the error in MASH arising from the violation of time-reversal symmetry.\cite{MASH} This arises due to the presence of a weighting factor in the definition of the MASH correlation functions that is not conserved by the dynamics.

For the specific case of the adiabatic population difference studied in this work, the MRE is defined by
\begin{equation}
\left\langle\left[|S^{\mathrm{ad}}_z(t)|-|S^{\mathrm{ad}}_z|\right]\sgn(S^{\mathrm{ad}}_{z}(t)) \right\rangle_0, 
\end{equation}
where $\tfrac{1}{2}$ and $\sgn(S_{z}^{\mr{ad}})$ are the MASH representations of the $\tfrac{1}{2}\hat{\mathcal{I}}_{\mr{s}}$ and $\hat{\sigma}_{z}^{\mr{ad}}(x)$ operators respectively, $2|S_{z}^{\mathrm{ad}}|$ is the corresponding MASH weighting factor for this correlation function, and the ensemble average $\braket{\cdots}_{0}$ is defined in \cref{eq:Ai_ave}.

Our formula for evaluating the long-time limit of quasiclassical correlation functions [\cref{eq:long_time_gen}] can be applied here to evaluate the long-time limit of the MRE\@. Using 
\begin{equation}
\int_{-1}^{1}\rd S_{z}^{\mr{ad}} \,\left\langle \left[S^{\mathrm{ad}}_z - \sgn(S^{\mathrm{ad}}_z)\right] \rho^{\mathrm{MASH}}_{\mathrm{eq}}(S^{\mathrm{ad}}_{z})\right\rangle_{\mr b}
= 0 \label{eq:M_MASH}
\end{equation}
we find that the MRE vanishes in the long-time limit. To obtain this result, we have also used $\langle |S^{\mathrm{ad}}_z|\rangle_0=1$ and the definitions of the the ensemble average $\braket{\cdots}_{\mr{b}}$ and the MASH thermal density given in \cref{eq:bath_int,eq:MASH_dist} respectively. This result therefore explains why MASH is able to exactly reproduce the long-time limit of correlation functions despite not rigorously obeying detailed balance. 

\end{appendix}
\FloatBarrier
%


\end{document}